\documentclass[iop]{emulateapj}
\usepackage{amsmath}
\usepackage{mathtools}
\usepackage{epstopdf}
\def\spose#1{\hbox to 0pt{#1\hss}}
\def\simlt{\mathrel{\spose{\lower 3pt\hbox{$\mathchar"218$}}
     \raise 2.0pt\hbox{$\mathchar"13C$}}}
\def\simgt{\mathrel{\spose{\lower 3pt\hbox{$\mathchar"218$}}
     \raise 2.0pt\hbox{$\mathchar"13E$}}}

\shorttitle{M31 Age-Dispersion Relation}
\shortauthors{Dorman et al.}
\slugcomment{Accepted for publication in the Astrophysical Journal}
\begin{document}
\bibliographystyle{plainnat}
%--------------------------------------------------------
%  TITLE & AUTHOR LIST
%--------------------------------------------------------
\title{
A clear age-velocity dispersion correlation in Andromeda's stellar disk
}
\author{Claire~E.\ Dorman\altaffilmark{1}, 
Puragra\ Guhathakurta\altaffilmark{1}, 
Anil~C.\ Seth\altaffilmark{2},
Daniel~R.\ Weisz\altaffilmark{3, 4},
Eric~F.\ Bell\altaffilmark{5},
Julianne J.\ Dalcanton\altaffilmark{3},
Karoline M.\ Gilbert\altaffilmark{6},
Katherine~M.\ Hamren\altaffilmark{1},
Alexia~R.\ Lewis\altaffilmark{3},
Evan~D.\ Skillman\altaffilmark{7},
Elisa~Toloba\altaffilmark{1},
Benjamin~F.\ Williams\altaffilmark{3}
}

\altaffiltext{1}{UCO/Lick Observatory, University of California
    at Santa Cruz, 1156 High Street, Santa Cruz, CA~95064; {\tt [cdorman,
      raja, khamren, toloba]@ucolick.org}}
\altaffiltext{2}{Department of Physics \& Astronomy, University of
  Utah, Salt Lake City, UT 84112, USA; {\tt aseth@astro.utah.edu}}
\altaffiltext{3}{Department of Astronomy, University of Washington,
  Box 351580, Seattle, WA 98195, USA; {\tt [dweisz,
    jd, arlewis, ben]@astro.washington.edu}}
 \altaffiltext{4}{Hubble Fellow}
  \altaffiltext{5}{Department of Astronomy, University of Michigan, 500 Church Street, Ann Arbor, MI 48109, USA; {\tt ericbell@umich.edu}}
\altaffiltext{6}{Space Telescope Science Institute, Baltimore, MD 21218, USA; {\tt kgilbert@stsci.edu}}
  \altaffiltext{7}{Minnesota Institute for Astrophysics, University of Minnesota, Minneapolis, MN 55455, USA; {\tt skillman@astro.umn.edu}}
%--------------------------------------------------------
 %  ABSTRACT
%--------------------------------------------------------
\begin{abstract} 

The stellar kinematics of galactic disks are key to constraining disk formation and evolution processes. In this paper, for the first time, we measure the stellar age-velocity dispersion correlation in  the inner 20
kpc ($\sim 3.5$ disk scale lengths) of M31 and show that it is dramatically different from that in the Milky Way. We use optical Hubble Space Telescope/Advanced Camera for Surveys 
photometry of $5800$ individual stars from the Panchromatic Hubble Andromeda Treasury (PHAT)
survey and Keck/DEIMOS radial velocity measurements of the same stars from the Spectroscopic and Photometric Landscape of Andromeda's Stellar Halo (SPLASH) survey. We show that the average line-of-sight velocity dispersion is a steadily increasing function of stellar age exterior to $R=10~\rm kpc$, increasing from $30~\rm km~s^{-1}$ for the main sequence stars to $90~\rm km~s^{-1}$ for the red giant branch stars. This monotonic increase implies that a continuous or recurring process contributed to the evolution of the disk. Both the slope and normalization of the dispersion vs. age relation are significantly larger than in the Milky Way,
allowing for the possibility that the disk of M31 has had a more violent history than
the disk of the Milky Way, more in line with $\Lambda$CDM predictions. We also find evidence for an inhomogeneous distribution of stars from a second kinematical component in addition to the dominant disk component. One of the largest and hottest high-dispersion patches is present in all age bins, and may be the signature of the end of the long bar. 
\end{abstract}
 
 %-------------------------------------------------------
 %  KEYWORDS
 %-------------------------------------------------------
%--------------------------------------------------------
 %  1. INTRODUCTION
 %-------------------------------------------------------
\section{INTRODUCTION}\label{intro_sec}

Most late-type galaxies have multiple disk populations with distinct
scale heights. 
%Thick disks appear ubiquitous in late-type galaxies in the the local
%universe.
The vertical surface brightness profiles of edge-on
spirals are well fit by double exponential profiles \citep{yoa06}, while dynamically
hot ``thick disk'' populations have been found in both
the Milky Way \citep[e.g.,][]{chi00} and Andromeda \citep{col11} galaxies. While it is unclear whether galactic disks
typically contain two distinct components or a continuum of progressively
thicker populations as argued in \citet{bov12}, the structure of the
stellar disk is an important key to understanding a galaxy's formation history. 

Thick disks are thought to have formed through some combination of the
following three processes. First,  stars can be formed in a cold, thin disk and later be dynamically heated by satellite impacts \citep{qui93, vel99, pur10, tis13} or by internal perturbers such as spiral arms, bars, or scattering by giant molecular clouds \citep{ida93}. Second, they can be formed ``in situ,'' from a
thick, clumpy gas disk at high redshift whose remnants collapse further over time
to form progressively younger, thinner, more metal-rich stellar disks \citep{bou09, for12}.
Third, the
dynamically hot population can be accreted from satellite
galaxies through tidal interactions \citep{aba03}. Each scenario should produce a different relationship
between age, metallicity, and degree of heating. For example, accretion of metal-poor satellites onto a thin disk would create a binary disk structure with thin and thick components, each with a distinct vertical scale height, age distribution and metallicity distribution. In contrast, a continuous process such as collapse of a clumpy gas disk or heating from frequent low-mass satellite impacts would produce disk layers whose thickness (degree of heating) increases with age.  

In nearby, low-mass, edge-on spiral galaxies, stellar populations' vertical scale
heights increase with age over three age bins, suggesting that a continuous
process such as disk heating plays a role in the evolution of those
galaxies \citep{set05}. An alternative heating diagnostic to scale height is velocity dispersion ($\sigma_v$).  Measuring $\sigma_v$ as a function of age can yield even stronger physical constraints on 
possible disk evolution mechanisms. Dispersion measurements are also less sensitive to dust than are scale height measurements, and are possible in galaxies that are not perfectly edge-on. 

However, meaningful kinematical measurements are difficult to make. 
In distant galaxies, kinematics derived from
integrated-light spectra cannot differentiate between the
contributions from old, intermediate-age, and young stellar
populations. In particular, near-infrared light from old red giant branch (RGB)
populations is contaminated by flux from younger asymptotic giant
branch (AGB) stars.  In the Milky Way itself,
we can more easily separate stars by age. Velocity dispersion appears
to increase monotonically with age in the solar neighborhood
\citep{nor04}, but the rest of the disk is obscured to the extent that
it is impossible to  tell if the solar neighborhood is representative of the entire disk, much less to trace large-scale kinematical structure across the Galaxy. 

Additionally, a study of only one disk galaxy (for example, the MW) does not allow us to draw general conclusions about the structure of disk galaxies in general. In the case of the MW, this is an important concern. $\Lambda$CDM cosmology predicts that galaxies are built up via accretion of smaller satellites. While most collisions between host and satellite occur in the halo of the host, a disk the mass of the MW's should have experienced at least one encounter with a massive ($\sim 3 M_{\rm disk}$) satellite \citep{ste08}. Such an encounter should significantly thicken and heat the disk \citep{pur09}, but the disk of the MW does not exhibit any such signs of a cosmologically common heating event. Studying the detailed structure of a second galaxy disk -- like that of Andromeda (M31) -- can provide an important constraint on whether $\Lambda$CDM needs to be revisited or whether the Milky Way is simply an outlier in the collision frequency distribution.

M31 is an ideal candidate for mapping a disk's velocity dispersion as a function of age. It is close enough ($785~\rm kpc$; \citet{mcc05}) that we can isolate and map
the velocity dispersions of the RGB, AGB, and young upper main sequence  (MS) populations
separately, but distant enough
that we can see the entire disk. We take advantage of data from two surveys of
the disk-dominated region of M31. The Spectroscopic
and Photometric Landscape of Andromeda's Stellar Halo (SPLASH) survey
has used the Keck/DEIMOS multiobject spectrograph to
measure radial velocities of thousands of individual bright stars in the inner $20$
kpc ($\sim 3.5$ disk scale lengths) of M31 \citep{gil09,dor12,dor13,how13}. Meanwhile, the
recently-completed Panchromatic Hubble Andromeda Treasury (PHAT)
survey, a Hubble Space Telescope MultiCycle Treasury (HST/MCT)
program, has obtained six-filter photometry of 117 million
individual stars in the same portion of the galaxy \citep{dal12, wil14}, allowing clean
color/magnitude-based separation of RGB, AGB, and MS stars.

Previously, we analyzed the kinematics of {\em only} the RGB stars in the intersection of the SPLASH and PHAT surveys. We found that $20\%$ of the RGB stars our survey belonged to a population wtih spheroid-like kinematics: with a velocity dispersion of $150~\rm km~s^{-1}$ and $v_{\rm rot}/\sigma \sim 1/3$ (the "inner spheroid"; \citet{dor12}). Later, we found that the inner spheroid population has a disk-like luminosity function despite its spheroid-like kinematics \citep{dor13}. In the current paper, we expand our survey to include three younger PHAT photometry-defined age bins in addition to the RGB population, and map the line-of-sight (LOS) velocity dispersion of stars in each age bin. In this paper, we aim to characterize the bulk properties of the disk in each age bin and so do not distinguish between ``disk" and ``spheroid" members. However, in \S\,\ref{sec_discussion}, we discuss the evidence for and possible origins of subsets of the RGB and AGB populations with atypical LOS velocities. 

This paper  is organized as follows. First, in \S\,\ref{sec_data}, we
present the dataset composed of Keck/DEIMOS radial velocity
measurements and HST optical photometry of over $8200$ individual
stars. In
\S\,\ref{sec_methods}, we separate the stars into four age
bins based on their position in the optical color-magnitude diagram
(CMD) and estimate RGB photometric metallicities. We also define our smoothed velocity dispersion statistic. We discuss
trends and possible biases in \S\,\ref{sec_results}. In
\S\,\ref{sec_discussion}, we discuss the constraints our results place
on  disk evolution scenarios. Finally, we summarize in
\S\,\ref{sec_summary}.

\section{Data acquisition and reduction}\label{sec_data}

\begin{figure*}
\scalebox{0.8}{\includegraphics[trim=10 0 30 20, clip =
  true]{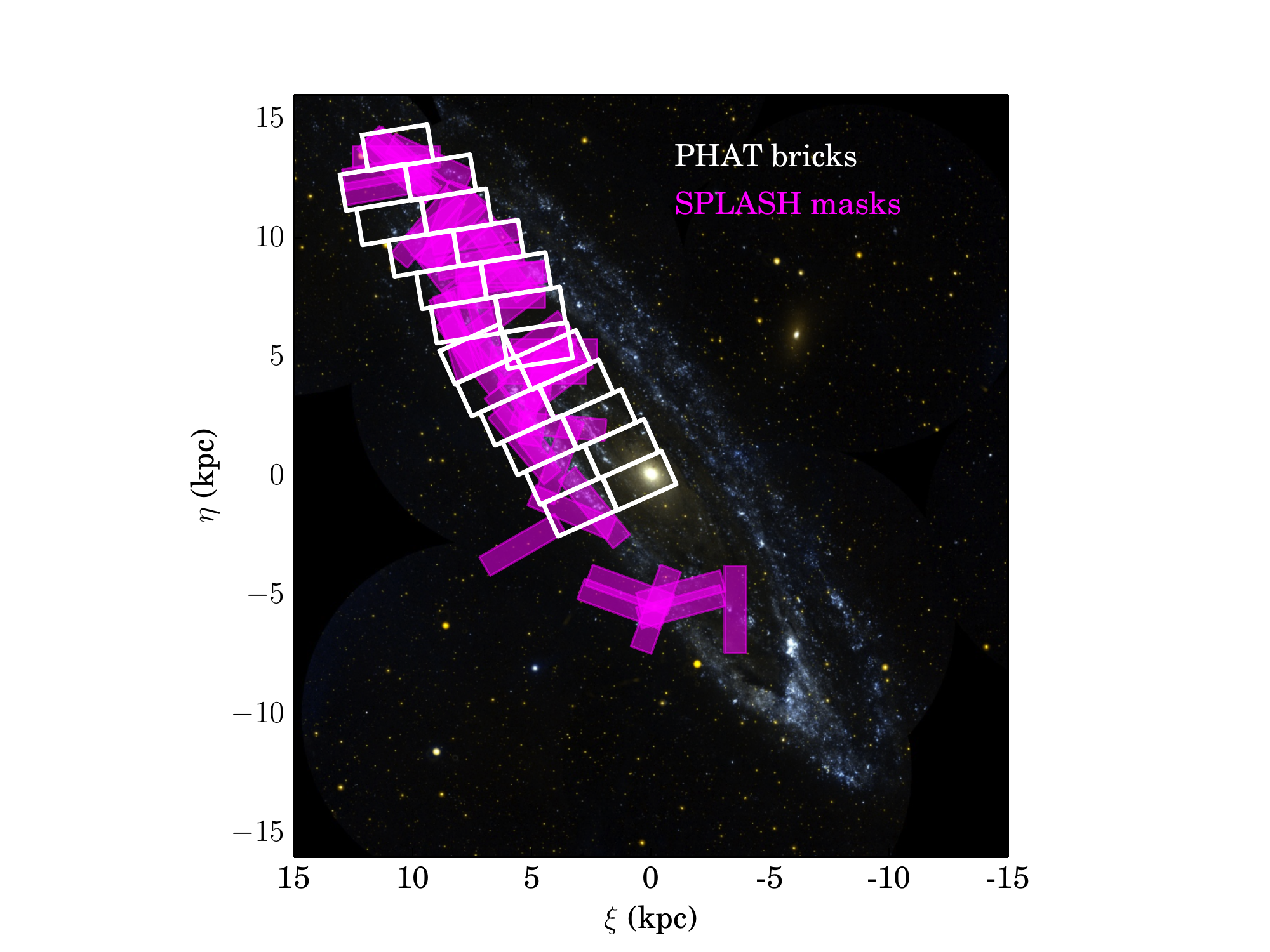}}
\centering
\caption{Spatial coverage of spectroscopic (SPLASH) and photometric (PHAT) surveys
  from which our data are drawn, overlaid on a GALEX UV image of
  M31 for reference.  Magenta regions demarcate the 47 Keck/DEIMOS spectroscopic slitmasks used in the SPLASH survey, whereas white rectangles outline the 23 PHAT ``bricks" (clusters of HST pointings). In this paper, we use only stars in the intersection of these two surveys: those with both PHAT optical photometry and reliable SPLASH-derived radial velocities.}\label{fig_surveymap}
\end{figure*}

Our dataset is a subset of two larger surveys. 
Figure~\ref{fig_surveymap} shows the spatial coverage of of both overlaid on a GALEX image of M31. We start with radial velocity measurements and optical
HST photometry of $8265$ stars in
the inner 20 kpc of M31, the region dominated by the visible disk. 
The radial velocities were measured using the Keck/DEIMOS multiobject
spectrograph between 2007 and 2012 as part of the SPLASH survey
\citep{gil07, dor12, dor13}. 
The photometry is from HST/ACS/WFC via
the PHAT survey \citep{dal12}.  The filters used were ACS F475W and F814W, which, for context, are roughly equivalent to Sloan $g$ and Cousins $I$ bands. Here we briefly describe the spectroscopic sample;
for information on the PHAT photometry, see \citet{dal12} and \citet{wil14}. 

This paper combines spectroscopic data from several smaller projects \citep{gil07, dor12, dor13}. Therefore, the spectroscopic target selection function is not homogeneous. $44\%$ of the
targets, those observed  between 2007 and 2010,  were selected based on their apparent degree of
isolation in a single-filter $i'$ CFHT/MegaCam mosaic image. These
targets primarily trace the dominant population in M31's
stellar disk: old, metal-rich RGB stars. Details on target selection
techniques for these masks can be found in \citet{dor12}. 

Because these objects are from masks designed prior to acquiring PHAT photometry at that location, they had to be later matched to their corresponding sources in the
six-filter PHAT photometric catalogs.  First, astrometric offsets
between each DEIMOS mask
and the PHAT coordinate system were obtained and applied to the spectroscopic sample. Then, for each object in the spectroscopic
sample, the brightest (in $m_{\rm F814W}$) PHAT star within a search radius of $0.''5$
was chosen as the match. 

The remaining $56\%$ of the spectroscopic targets used in this work, observed in
2011 and 2012, were chosen based on
existing PHAT photometry, eliminating the need for post-spectroscopy
cross-catalog matching. Since we had color information at the target selection stage, we were able
to prioritize under-represented populations over the dominant
metal-rich red giants. For the five slitmasks, targeting about 1000 sources,
observed in 2011, we chose red giants across the broad
range of photometric metallicities $-2.0 < [M/H] < 0.2$. We restricted our sample to stars that
were in the magnitude range $20 < m_{F814W}<22$, but otherwise chose
stars randomly in position and magnitude space.  (The quoted magnitudes, like all magnitudes in this paper, are in the Vega system.) See \citet{dor13} for
more information on the HST-aided spectroscopic target selection in 2011.

In 2012, we targeted about 4500 more stars across a broad range of
ages, including young massive MS stars, intermediate-age AGB stars, and RGB
stars. We also targeted a few young clusters identified by the PHAT
team \citep{joh12}, although those are not used in this work. All of our targets were brighter than either $m_{F814W}=22$ or
$m_{F475W}=24$. Faint RGB stars (with $m_{F814W}>21.5$) and faint MS
stars  (with $m_{F814W}>21$ or $m_{F475W}>23$) were given very
low priority and were only used on the rare occasion that there was unused space to fill on the masks, but otherwise targets were chosen randomly within each
evolutionary stage. These data are presented here for the first time;
see Appendix A for a full data table. 

\begin{figure*}
\scalebox{0.9}{\includegraphics[trim=15 0 20 20, clip =
  true]{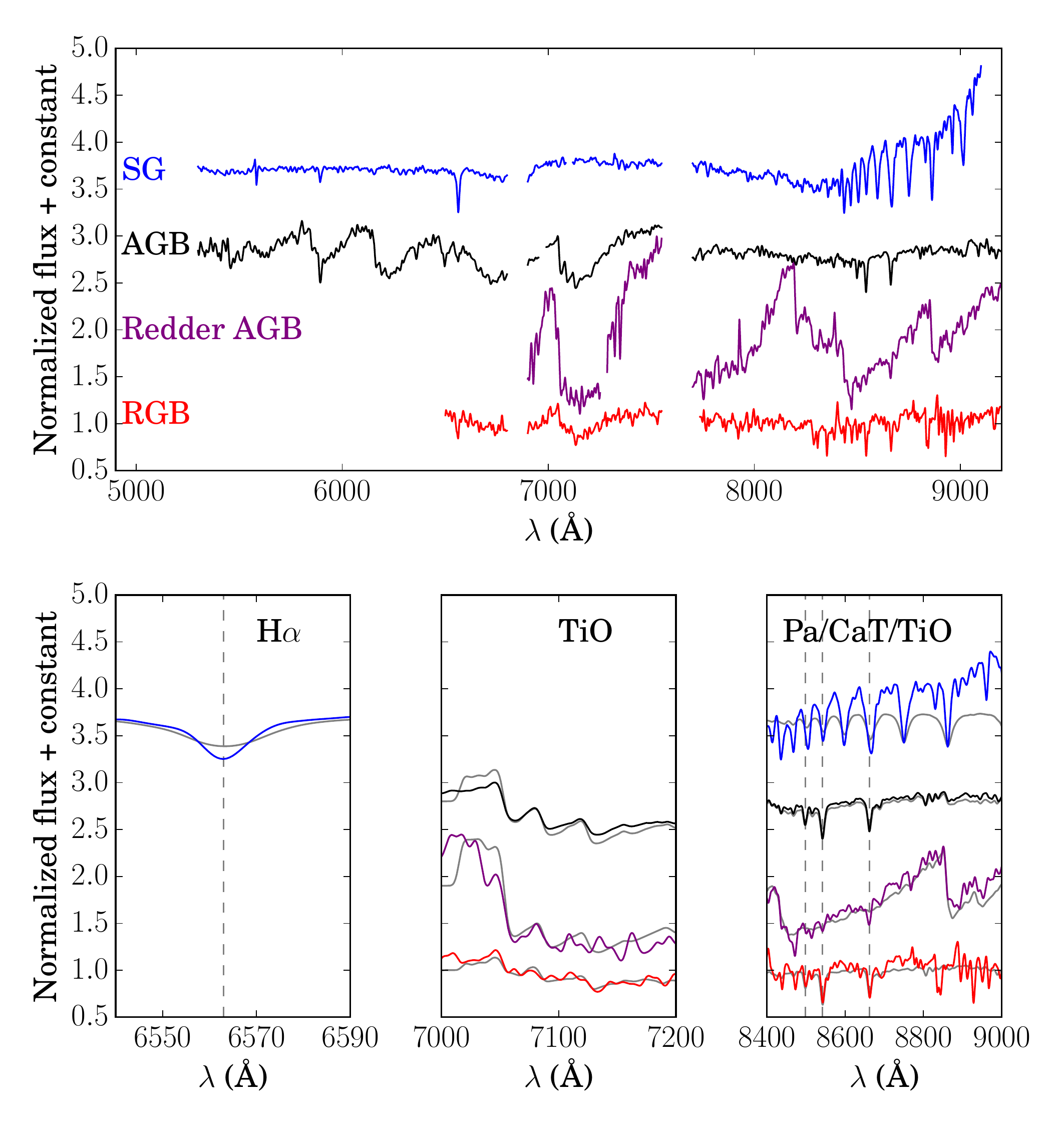}}
\centering
\caption{Example spectra from a variety of stellar types. Spectra have been normalized and shifted to rest frame. {\em Top:} Representative spectra from a young supergiant, young-intermediate age AGB, older and redder AGB, and RGB stars. The two bottom spectra are from very red stars, and so their flux on the blue half of the spectrum is very noisy and not shown. Gaps in the spectrum correspond to either the DEIMOS chip gap or the atmospheric A-band, whose positions vary from spectrum to spectrum in the rest frame of the star. {\em Bottom:} Zoom-in views of the portions of the spectrum useful for obtaining velocities. The science spectra are color-coded as in the top panel, while template rest-frame spectra used to measure radial velocities via cross-correlation are shown in gray. The wavelengths of H$\alpha$ and the Calcium II triplet are shown in gray vertical lines to aid the eye. (Note that the young (blue) star in the rightmost panel displays the Paschen series, not the Calcium triplet, although some of the lines fall at similar wavelengths.)}\label{fig_spectra}
\end{figure*}

After observing, each raw 2D spectrum is collapsed in the spatial direction and
cross-correlated against a suite of template rest-frame spectra to
measure its radial velocity. Figure~\ref{fig_spectra} shows some representative spectra and zoomed-in views of the absorption lines that dominate the radial velocity measurement. Though the entire spectrum is used in the cross-correlation, certain absorption lines are most important in determining the radial velocity. The Ca II triplet near $8500$\AA~is strong in RGB and many AGB stars. Temperature-sensitive TiO bands across the red side of the spectrum, including a strong triplet near $7050$\AA~and another band near $8850$\AA, determine the velocity for many of the redder AGB and RGB stars, including those without strong CaT.  For some of the blue MS stars, H$\alpha$ and H$\beta$ are the only reliable lines present, but the Paschen series is also present for young supergiants. After the automatic cross-correlation, each spectrum is inspected by eye, and only robust velocities (those based on at least
two strong spectral features) are used for kinematical analysis. About 1/3 of the MS stars do not pass the quality cut --- since hot stars have so few spectral features in the optical, it was harder to recover a robust velocity. 

The velocity precision varies based on the instrument
settings that were used. Through 2011, we used the $1200$ line/mm grating
on DEIMOS, resulting in an approximate wavelength range of
$6500-9100$\AA~and a spectral resolution of $R=6000$. In 2012, we
used the coarser 600 line/mm grating to gain spectral coverage as
blue as $4500$\AA~and better characterize the younger massive upper
MS stars, yielding a resolution of about $R=2000$. In both
cases, the width of each slitlet was $0.''8$. The radial
velocity uncertainties derived from the cross-correlation are on the
order of a few $\rm km~s^{-1}$ for the higher-resolution spectra, and $\sim 10~$
$\rm km~s^{-1}$ for the spectra from 2012. 

The stars in our full spectroscopic sample fall in the region of the galaxy dominated by the disk. Based on surface brightness profile decompositions \citep{cou11, dor12}, the bulge contributes only $2\%$ of the $I$ band surface brightness at the innermost portion of the survey and essentially zero exterior to about $8~\rm kpc$. Therefore, we do not need to remove bulge stars from the sample. 

%============================================

%============================================

\section{Methods: Velocity dispersion as a function of age and of
  metallicity}\label{sec_methods}

Our goal is to measure the line-of-sight (LOS) velocity dispersion of the stellar disk
as a function of age and of metallicity. In this section, we describe
our analysis procedure. First, we define regions in the optical CMD
corresponding to very young MS,
younger AGB, older AGB, and  old RGB populations using a simulated optical
CMD. Next we split our spectroscopic sample into those four bins using
the observed optical PHAT CMD. Then, we
map the line-of-sight dispersion of each component. Finally, to look
at the old population in more detail, we further split the RGB
population into two metallicity bins and construct a dispersion map
for the stars in each bin. 

\subsection{Definition of age bins}\label{sec_ages}

We first define four age bins in the optical F814W/F475W CMD using two
criteria: First, we use both photometric and spectral discriminants to identify stars that are unlikely to be MW foreground (MWFG) stars. Second, we use a simple simulated CMD to identify regions
containing stars of similar ages. We then roughly estimate the average ages
of our four age bins using the simulated CMD and point out that the age estimates of the older bins depend significantly on the assumed star formation history.

%the TRILEGAL model of
%the Milky Way \citep{van09} to identify regions in the CMD that are minimally
%affected by foreground contamination. }

\subsubsection{Foreground contamination}

MWFG dwarfs can lie in the same magnitude window as our M31
spectroscopic targets. We took steps, both pre- and post-spectroscopy,
to eliminate them from the catalog. 

%XXX Show a plot of this with the TRILEGAL model? Need to cite Leo 
For PHAT-selected targets, we avoided likely MW members in the target
selection stage using UV-IR color-color cuts. Here the UV color is ACS $F336W-F475W$ and the IR color is $F110W-F160W$. A comparison of the MW foreground as simulated by the TRILEGAL galaxy model \citep{van09} and a toy model of M31 shows that the foreground dwarfs are exclusively bluer in the IR color and redder in the UV color than the M31 giant sequence for stars with $F475W - F814W > 2$ --- that is, all of the RGB and AGB stars in our M31 sample. We use the TRILEGAL MWFG simulation to define a box in color-color space to exclude from our spectroscopic target selection. 

\begin{figure*}
\scalebox{0.77}{\includegraphics[trim=0 20 0 10, clip = true]{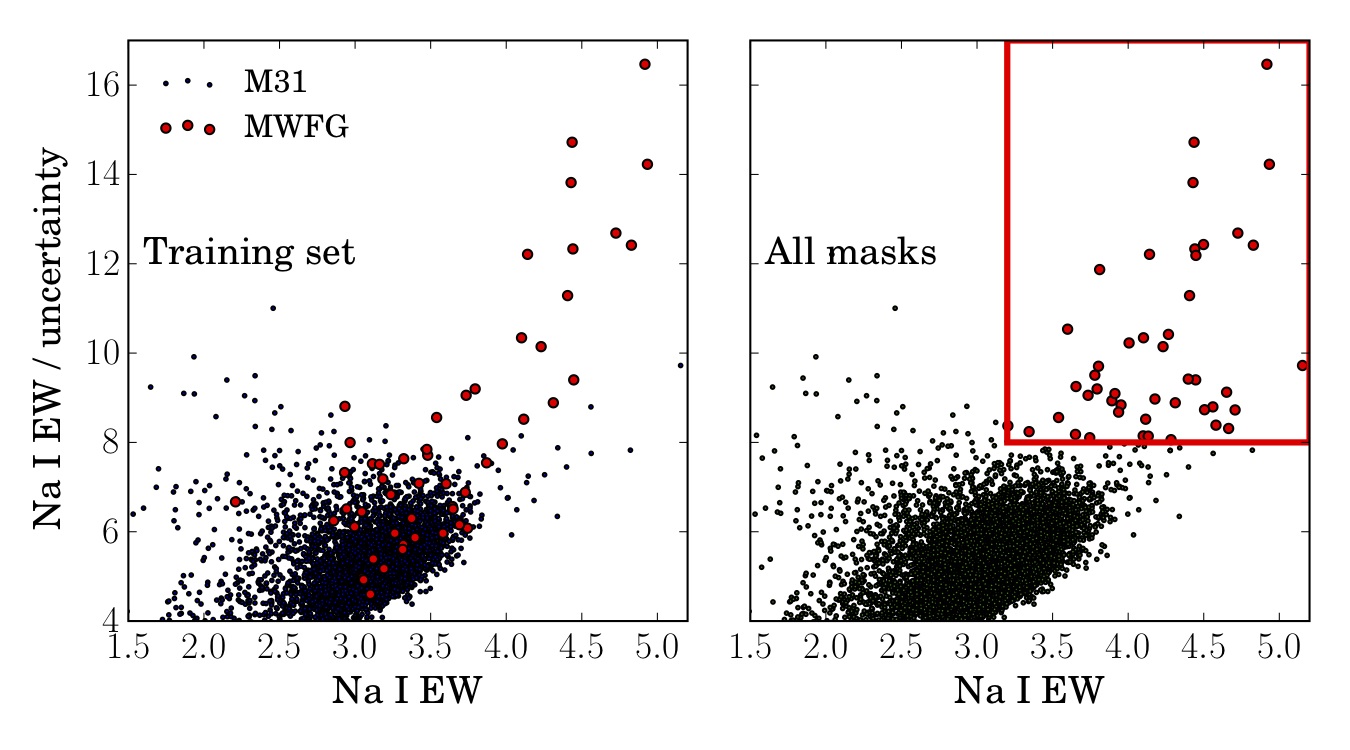}}
\centering
\caption{Identifying and removing likely MW foreground dwarfs based on
the presence of the surface gravity-sensitive Na I doublet. {\em
  Left:} training set based on about 8000 stars that were visually
checked for the presence of the doublet. Stars with equivalent widths
(EW) greater than 3.2 and significance (EW / uncertainty on EW)
greater than 8 are were almost universally flagged as dwarfs. {\em Right:}
Same discriminant applied to the full set of spectra (not all of which
had been manually inspected). Stars within the red box are almost
certainly MW dwarfs and were excluded from the analysis.}
\label{fig_na}
\end{figure*}

%excluding stars from the analysis after spectroscopy via the Na I
%doublet; look up references from Raja and add a plot
For the CFHT-selected targets, color information from PHAT was not available as a tool for excluding MWFG stars at the target selectino stage. Instead, we
identified red foreground dwarfs based on the presence of the Na I doublet at $8190$ \AA, which is an useful giant / dwarf discriminant due to to its sensitivity on surface gravity and temperature \citep{sch97}.  %XXX CITATION
We visually inspected each spectrum taken between 2007 and 2011, and
about half of the spectra taken in 2012, for the presence of the Na I doublet. We also
calculated the equivalent width (EW) and the uncertainty on the EW
measurement ($\sigma_{\rm EW}$) across the doublet bandpass (8179-8200 \AA) relative to the adjacent continuum (8130-8175\AA~and 8210-8220\AA) as in \citet{gil06}. 
The set of stars with both EW measurements and visual flagging formed
a training set from which we found a diagnostic that can be used to
automatically identify definite foreground dwarfs. Figure~\ref{fig_na}
shows the training set (left) and all stars with spectra
(right). Stars with EW$> 3.2$ and significance EW/$\sigma_{\rm EW} > 8$ (those
inside the red box) are foreground dwarfs and were eliminated from the
spectroscopic sample. This discriminant was applied to all stars in the sample, including PHAT-selected targets that had already survived the pre-spectroscopy color-color cut. 

%quantifying the effect of remaining contamination via the TRILEGAL
%MW model 
The steps described above do not necessarily eliminate every
foreground star. To esimate the number of contaminants that remain in
the sample as a function of optical color and magnitude, we employ the
TRILEGAL simulation of the Milky Way in the direction of M31. In each
color/magnitude bin (CM), we calculate the number of contaminants
$N_{\rm MW, expected, CM}$ expected
in the SPLASH survey: 

\begin{eqnarray}
N_{\rm MW, expected, CM} = N_{\rm SPLASH, CM}\frac{N_{\rm Trilegal, CM}}{N_{\rm PHAT, CM}}
\end{eqnarray}

We then subtract the number of contaminants already identified in and
removed from  that
color/magnitude bin based on the Na I doublet
discriminant: 

\begin{eqnarray}
N_{\rm contaminants, CM} = N_{\rm MW, expected, CM} - N_{\rm removed,
  CM}
\end{eqnarray}

\begin{figure}
\scalebox{0.8}{\includegraphics[trim=15 15 0 380, clip = true]{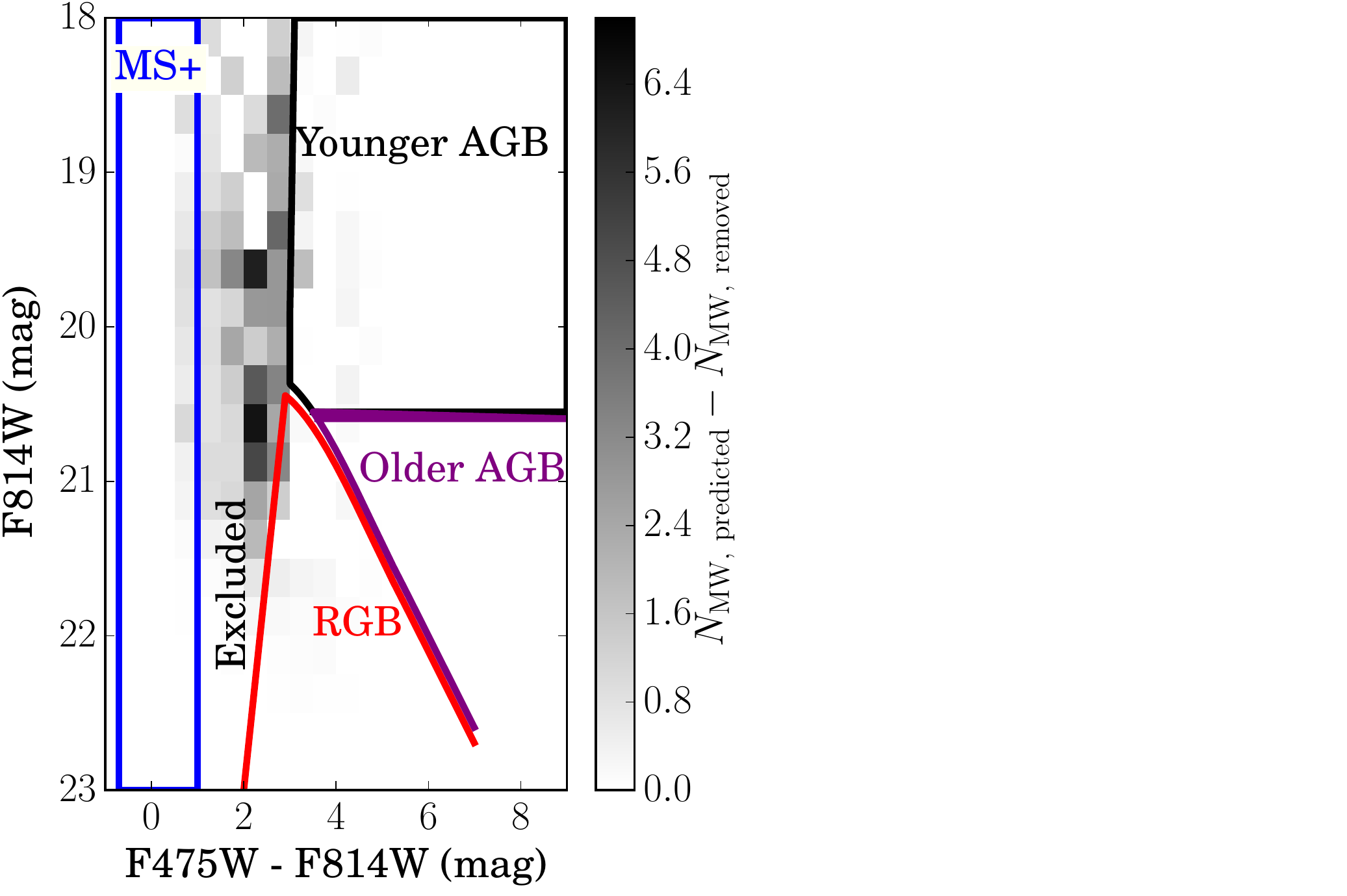}}
\centering
\caption{Optical Hess diagram showing the number of foreground
  Milky Way stars expected in our spectroscopic sample after cutting
  out stars with strong surface gravity-sensitive $Na I$ doublets,
  which are very likely to be foreground MW dwarfs. The blue, black,
  violet, and red lines outline the MS, younger AGB, older AGB, and
  RGB regions described and defined in \S\,\ref{sssec_simCMD}
  below. The most-contaminated portion of the CMD -- brighter than $m_{\rm
    F814W}=21$ and between $1 < m_{\rm F475W}-m_{\rm F814W} < 2$ -- is
excluded from the analysis.}
\label{fig_mwfg}
\end{figure}

 Figure~\ref{fig_mwfg} shows the resulting distribution of $N_{\rm
   contaminants, CM}$. The number is only significant ($>10\%$ of the
 sample) brighter than $m_{\rm
    F814W}=21$ and between $1 < m_{\rm F475W}-m_{\rm F814W} <
  2$. Stars in this portion of the CMD are excluded from the
  analysis. After exclusion, fewer than 0.1\% of the stars in the sample are expected to be MWFG stars.

\subsubsection{Estimating ages using a simulated CMD}\label{sssec_simCMD}

To estimate the ages of stars as a function of CMD position, we
simulate a simple stellar population in the optical CMD. 

Simulating a CMD requires choosing an age-metallicity relation (AMR) and star formation rate (SFR). A few measurements from the outer regions of M31's disk $(R\sim 20-30~\rm kpc)$ using deep HST photometry are available \citep{bro06, ber12}. Both show a clear inverse relationship between stellar age and metallicity. For our CMD, we are more interested in separating the CMD into regions with distinct average ages than pinning down those ages precisely. We assume a constant SFR of $1~M_{\odot}~{\rm yr}^{-1}$.  This assumption is overly simplistic --- for example, \citet{ber12} measure a burst of star formation in the outer disk at around 2 Gyr ago --- but is acceptable as we only aim to separate our CMD into age bins, rather than measure the precise ages of stars in those bins. We estimate the AMR empirically from the PHAT RGB data in the following way: we use 10 Gyr old PARSEC 1.1 \citep{bre12} isochrones in the metallicity range $-2.1\le [M/H] \le 0.3$
to estimate the metallicity distribution function (MDF) of all of the bright
($m_{\rm F814W} < 23$) RGB stars in the PHAT survey. The AMR is then
constructed so as to replicate the MDF, assuming a constant star
formation rate and a metallicity that never decreases with time. For example,
since there are $7$ times as many stars at
$[M/H]=-0.5$ as at $[M/H] = -1$, the simulation is allowed to spend
$7$ times as long producing stars with $[M/H]=-0.5$ as with $[M/H] =
-1$.  We use \citet{gir10} isochrones, a Kroupa IMF, assume that $35\%$ of
stars are in binaries, and apply a constant foreground reddening of
$A_v=0.2$.  

While this technique ignores the age-metallicity degeneracy on
the RGB, it generates a reasonable CMD. The CMD generated using this AMR is very similar to one produced using an AMR adapted from the empirical one presented in \citet{bro06} which used deep HST photometry of a small field in M31's disk about $25~\rm kpc$ from the galactic center.  Our CMD assumes a constant SFH and thus is not a SFH fit. Due to the age-metallicity-extinction degeneracy along the RGB, the assumed SFH may differ dramatically from the true one. However, it enables us to select four CMD regions with different average ages and measure approximate values for those average ages.

\begin{figure*}
\scalebox{0.8}{\includegraphics[trim=42 00 0 30, clip = true]{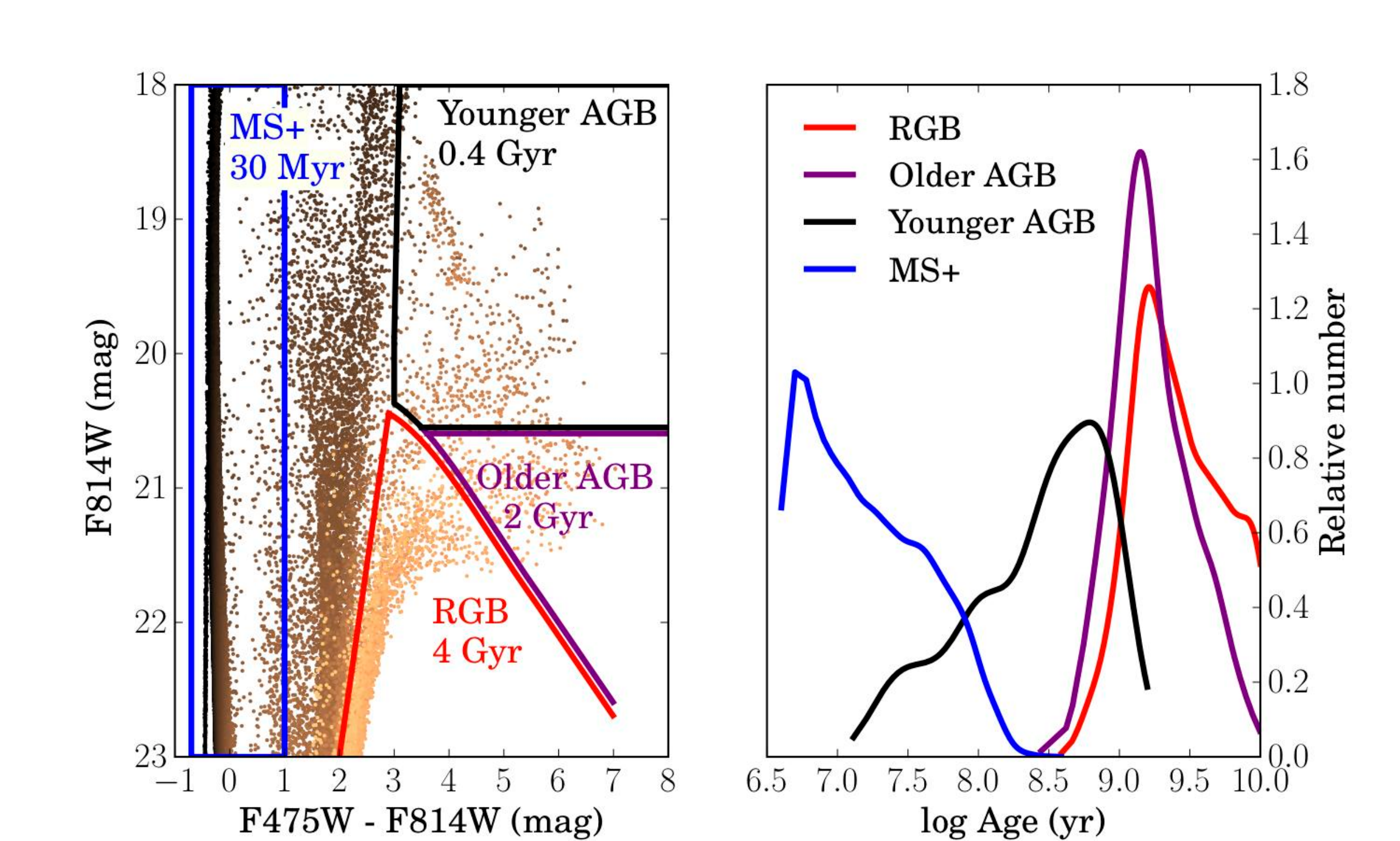}}
\centering
\caption{{\em Left:} Simulated CMD assuming a constant star formation rate. The blue,
  black, violet, and red outlines define the ``MS,'' ``younger AGB,''
  ``older AGB,'' and ``RGB'' age bins, respectively. Stars are color coded by
  log(age), with brighter (yellower) colors corresponding to older ages. Numbers correspond to the average age of stars within each bin. {\em Right:}
Age distributions for the same four regions. Age bin colors are the same as the left
panel and throughout this paper.}
\label{fig_sim}
\end{figure*}

We use the simulated CMD to define four regions containing stars of
increasing average age, while avoiding the highly contaminated region
described in the previous section. For simplicity, we refer to
these bins as ``MS+,'' ``younger AGB,'' ``older AGB,'' and ``RGB,''
although stars in a given bin do not necessarily all belong to the
exact same evolutionary stage. The constant-SFR CMD and age bin outlines are shown
in the left hand panel of Figure~\ref{fig_sim}; in the rest of this
section, we explain the choice of bins. 

All stars bluewards of $m_{F475W}-m_{F814W}=1$ are
classified as ``MS+'' stars. These bright blue stars are primarily
massive upper MS members younger than $100$ Myr,
although some may be blue supergiant (BSG) stars, which have similar
ages. 
The RGB region includes stars redwards of the line that passes through $(m_{\rm
  F475W}-m_{\rm F814W}, m_{\rm F814W})=(2, 23) {\rm~and~} (2.7, 20.4)$ and fainter than the tip of the red giant branch
(TRGB). The blue limit is chosen by eye so as to minimize contamination with
young MS stars, which overlap the oldest RGB stars in high-metallicity
systems such as M31. We measure the TRGB using the \citet{bre12}
isochrones described earlier. The brightest RGB stars on these
isochrones trace the TRGB as a function of $F475W-F814W$ color. 

We classify most of the rest of the red side of the CMD as AGB. To avoid
contaminating the AGB bins with older
RGB stars, we do not use stars
less than $0.1$ magnitude brighter than the TRGB (those within
photometric scatter of the TRGB).  

Stars of a large range of ages, from a few hundred Myr to
several Gyr, can lie on the AGB. At a given metallicity, younger
AGB isochrones are brighter and bluer than older AGB tracks. In the
simulated CMD, age roughly tracks luminosity, with younger stars
towards the top of the CMD. We use this age-luminosity dependence
to split the AGB in half (along the line $m_{\rm
  F814W} = 20.5$) into two age bins: ``younger AGB'' and ``older
AGB.''

The final classification scheme into four age bins is shown in the left-hand panel of
Figure~\ref{fig_sim}. The MS+ bin is outlined in blue on the blue side of
the optical CMD, RGB stars in red below the TRGB, older AGB stars in
purple brightward of the TRGB, and younger AGB stars in
black. The right hand panel shows the age distributions of stars in
these bins. The age distributions are broad and overlap, but have increasing mean ages: The MS+ bin has a mean age of $30~\rm Myr$, while the younger
AGB, older AGB, and RGB stars have average ages of $0.4$, $2$, and
$4$ Gyr, respectively. The RGB bin has a low average age for two reasons: First, because we imposed an age-metallicity relation, the older, bluer  metal-poor RGB stars actually overlap in CMD space with young red Helium burning stars, while younger, metal-rich RGB stars do not suffer from this ambiguity. Therefore we use only the younger (redder) portion of the RGB in this work. 
Second, and more importantly, the RGB age distribution in any magnitude limited sample with a constant SFR is biased towards younger ages since the rate of stars moving off the main sequence onto the red giant branch is higher for younger stars.

For comparison, we also generate a CMD with an exponentially decreasing SFR with timescale $\tau=4$ Gyr. This SFR is much steeper (skewed towards older stars) than seen in the outer disk of M31 \citep{ber12}, but gives an interesting boundary case. The boundaries of the most reasonable four age bins in CMD space are the same as for the constant-SFR simulation, but the average ages of the older age bins increase: the older AGB has an  average age of $3.5~\rm Gyr$ and the RGB has an average age of $5.5~\rm Gyr$. We do not show this CMD here since it does not influence our choice of age bin boundaries, but we discuss both sets of age estimates later in the paper.

%============================================

\subsection{Separation of data into age bins}\label{ssec_ages}

We divide the stars in the kinematical sample into the four age bins
defined in the previous section, using optical PHAT photometry. 

\begin{figure*}
\scalebox{0.8}{\includegraphics[trim=35 0 10 30, clip =
  true]{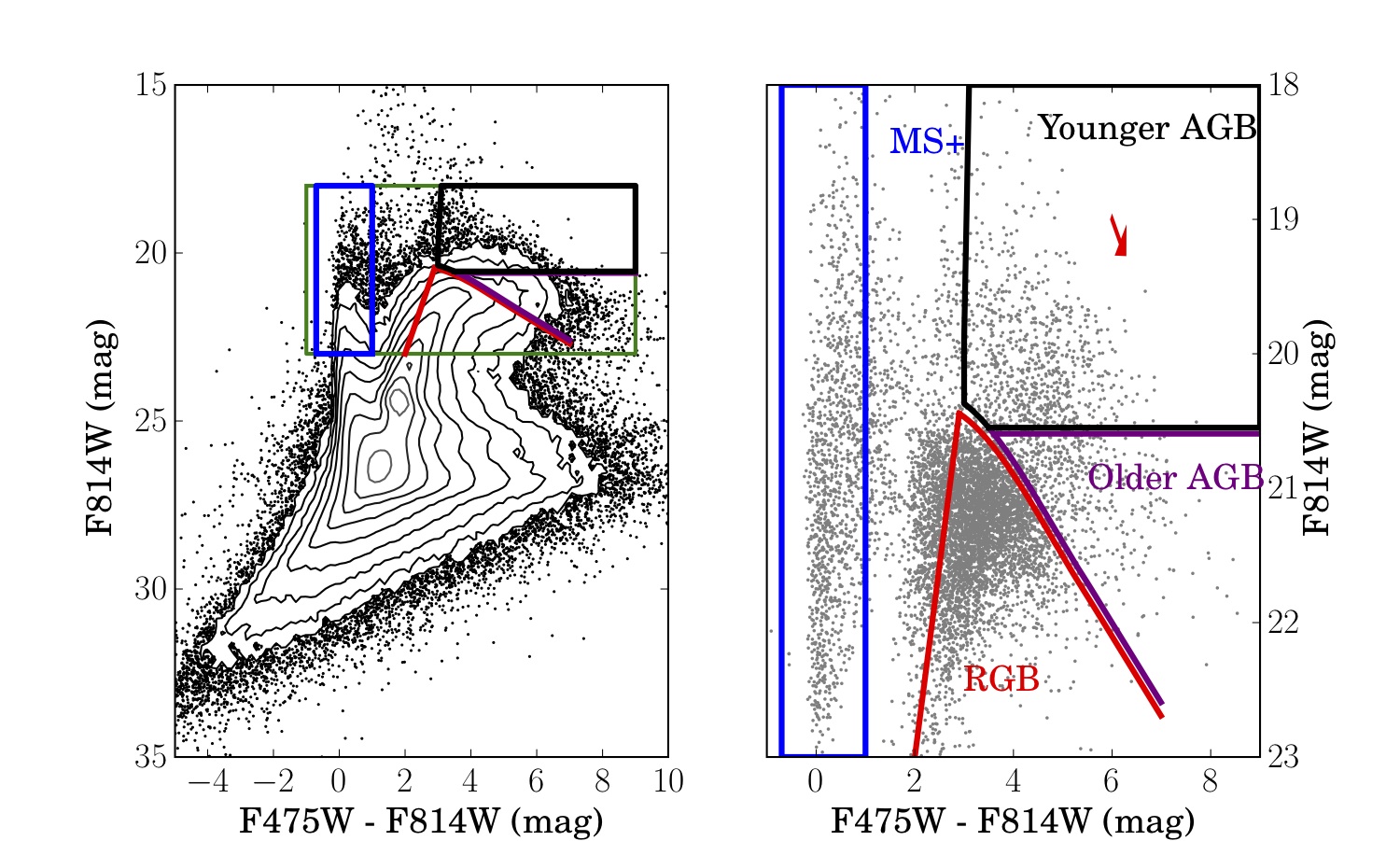}}
\centering
\caption{{\em Left:} Optical
 PHAT CMD from Brick 12, a representative brick near the middle of the PHAT survey area. The green box marks the portion of the CMD sampled by the
 spectroscopic survey. The age bin region outlines are the same as in Figure~\ref{fig_sim}.  {\em Right: }  Optical PHAT CMD of spectroscopic targets in the PHAT survey region that have optical PHAT photometry and reliable velocities and are unlikely to be foreground MW stars based on the strength of their Na I doublet lines. The age bin region outlines are the same as in the left hand panel and in Figure~\ref{fig_sim}.  Stars that fall outside the age bin boundaries have ambiguous ages and/or may be foreground stars, are not used in this work. The red arrow shows the median reddening vector due to dust in M31's disk, as measured by Dalcanton et~al. (2014), assuming that half of the spectroscopic targets lie behind M31's dust layer.
 Photometric errors are typically less than $0.01$ mag in each filter \citep{dal12}.}
\label{fig_cmd}
\end{figure*}

The left hand panel of Figure~\ref{fig_cmd} shows an optical Hess diagram of all stars in the PHAT survey in a representative region of the galaxy.  Spectroscopic targets fall into the color/magnitude range outlined in green. The right hand panel of Figure~\ref{fig_cmd} shows a CMD of the spectroscopic sample only, divided into age bins as in Figure~\ref{fig_sim}.  

While our simulated CMD does include reddening from the MW foreground, it does not account for differential extinction by dust within the disk, and we do not attempt to account for this shortcoming in classifying the data into age bins. This means that some of the CMD regions may be contaminated by stars from bluer age bins. However, the direction of the reddening vector is
such that only two bins are likely to be contaminated: A few younger AGB stars may be reddened into the older AGB region. The
RGB region is largely immune from contamination due to reddening, since the shape of the TRGB is such that old AGB stars will never be reddened into the RGB region, and the other two bins are
far enough away in CMD space that their members will not be reddened
onto the RGB either. Additionally, the red helium burning stars, which are not included in any of our bins, can be reddened into the RGB bin; however, these are much smaller in number than the RGB stars and thus largely insignificant even with reddening. However, the region in the CMD occupied by RGB stars contains stars of a range of ages, as discussed earlier and pictured in Figure~\ref{fig_sim}.

We estimate the average reddening vector in our sample using the M31 dust map presented in Dalcanton et~al. (2014). This map, constructed by comparing the infrared colors of RGB stars in the PHAT survey to the unreddened RGB, gives the average $A_V$ as a function of location across the disk.   Only stars that lie behind the dust layer are affected. The average reddening vector at the locations of our spectroscopic targets, assuming half the targets lie behind the dust layer, is shown in Figure~\ref{fig_cmd}. 

%============================================
%============================================

\subsection{Velocity dispersion maps}\label{ssec_sv}

\begin{figure*}
\scalebox{0.8}{\includegraphics[trim=20 30 10 25, clip =
  true]{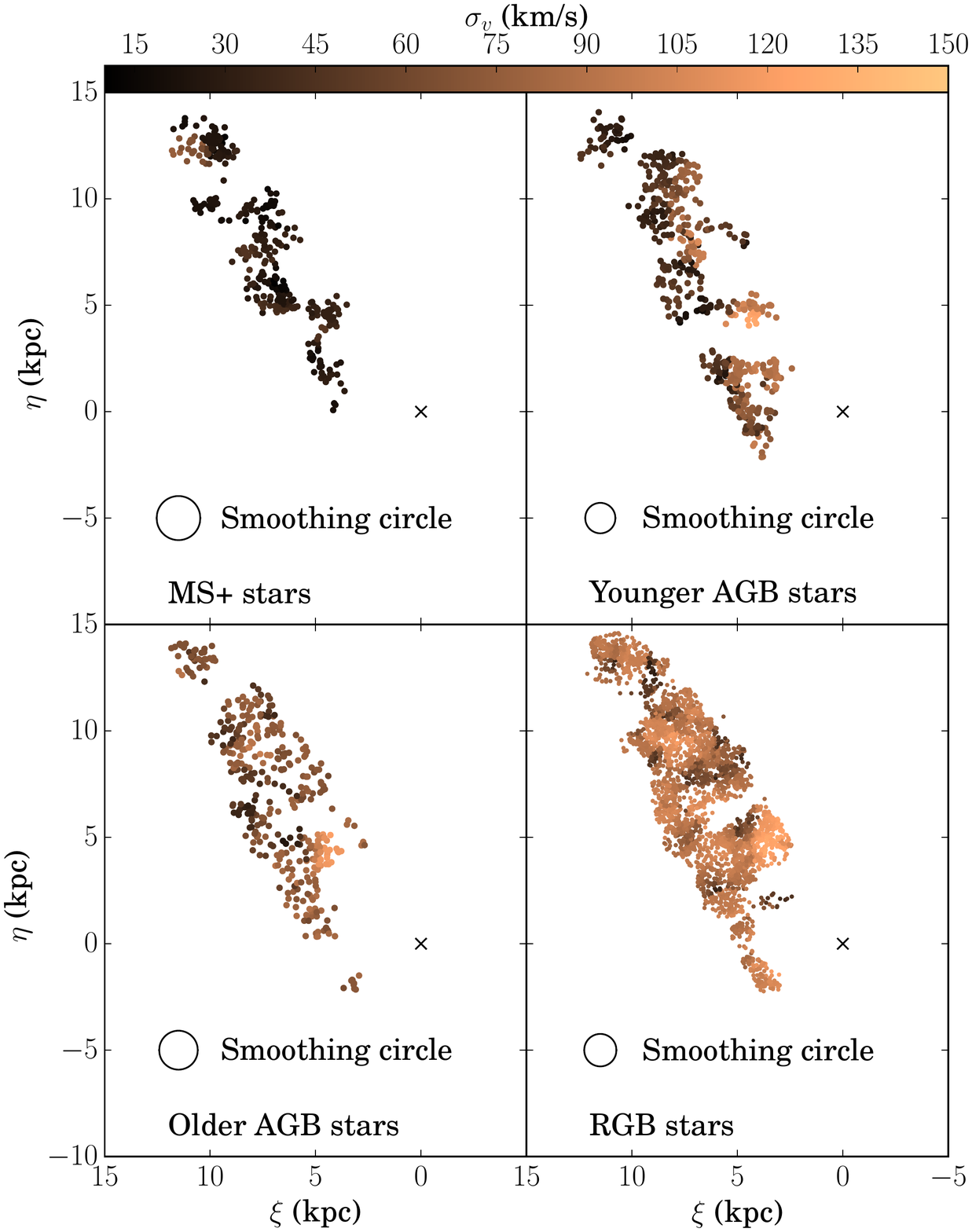}}
\centering
\caption{Smoothed, local, line-of-sight velocity dispersion of stars
  of different evolutionary stages: young main-sequence {\em (Upper
    Left)}; younger AGB {\em (Upper Right)}; older AGB {(\em Lower
    Left)}; and RGB {(\em Lower Right)}. The circles show the sizes of the smoothing circles in which the dispersion is calculated; smaller circles can be used for populations with higher number density. The typical dispersion increases with average age. The dispersion varies across the face of the disk within each age bin in a way that can be explained by our finite sampling density, as described in the Discussion section. \label{fig_sigmavmap}}
\end{figure*}

We now map the LOS velocity dispersion of stars in each of the four
age bins using a smoothing technique and display the result in two
ways: as a 2-D sky map (Figure~\ref{fig_sigmavmap}) and as a 1-D
dispersion distribution (Figure~\ref{fig_sighist}). We do not fit
separately for the radial, azimuthal and vertical components of the
dispersion. Such a decomposition requires assumptions on the rotation
curve and the geometry of the disk, and we choose to keep our analysis
purely empirical. A future paper will examine the shape of the
velocity ellipsoid. For reference, because of M31's nearly edge-on
inclination, the vertical component of the velocity dispersion has
negligible contribution to the LOS dispersion anywhere on the
disk. In general, the LOS dispersion is a combination of the radial
and azimuthal dispersion components, though the azimuthal component
dominates near the major axis where most of the survey field lies.

 For each target, we measure
the weighted second moment of the velocity distribution of all neighbors that belong to the same age bin and also fall within some radius of that target using the maximum
likelihood method described in \citet{pry93}. The weights are the inverse square of the velocity measurement uncertainty. This technique takes
into account random scatter from individual velocity measurement
uncertainties, and also allows
a straightforward computation of the uncertainty on the velocity
dispersion estimate as long as the number of points is at least $\sim 15-20$. 

\begin{figure*}
\scalebox{0.85}{\includegraphics[trim = 5 35 40 350, clip =
  true]{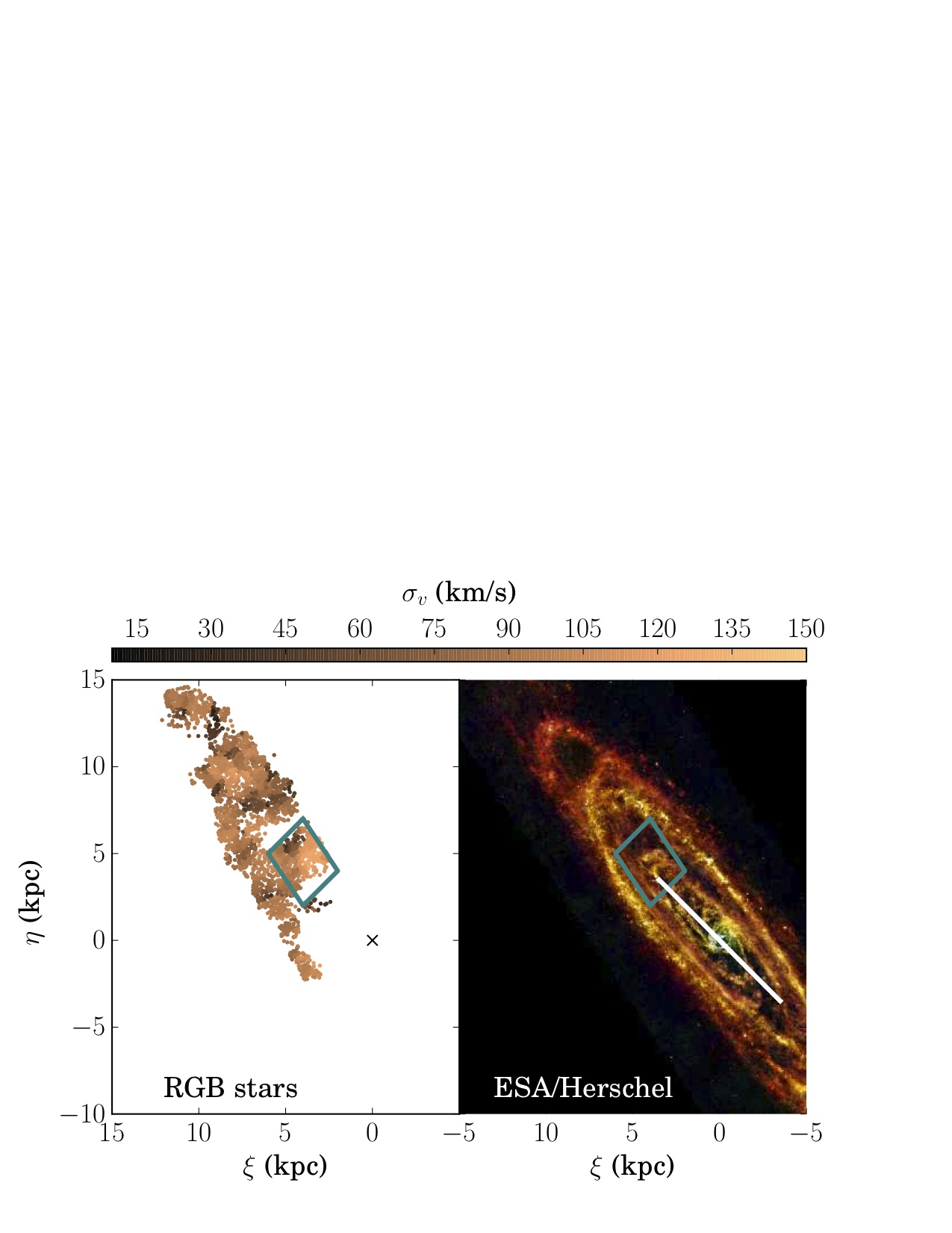}}
\centering
\caption{RGB dispersion map from Figure~\ref{fig_sigmavmap} next to a
  Herschel image of M31 for reference. A high-dispersion region which we term the ``Brick 9 Region" is outlined in teal in each panel.  The Brick 9 region aligns with the end of the inner ring, also roughly consistent with the end of the long bar according to the simulations by \citet{ath06}. The estimated size and orientation of the bar from \citet{ath06} is marked with a white line.  \label{fig_herschel}}
\end{figure*}

\begin{figure}
\scalebox{0.85}{\includegraphics[trim=20 40 200 70, clip =
  true]{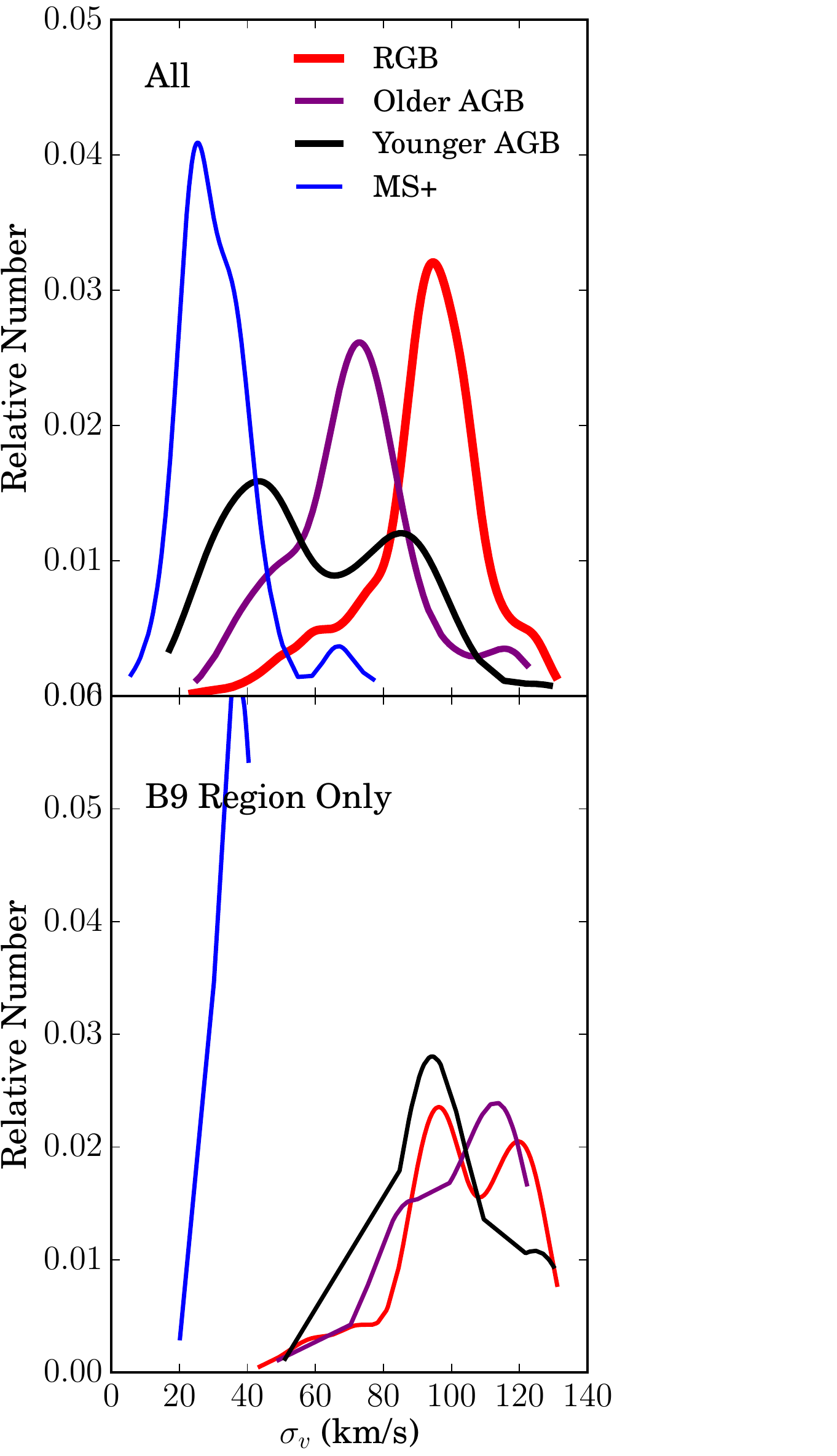}}
\centering
\caption{Local line-of-sight velocity dispersion distributions of stars in the four age bins for the entire survey area {\em Top} and only stars in the Brick 9 region outlined in Figure~\ref{fig_herschel} {\em Bottom}. For this plot, smoothing circle sizes were chosen independently for each point such that the uncertainty in velocity dispersion was $7, 10, {\rm and} 5~\rm km~s^{-1}$ for the RGB, AGB, and MS age bins, respectively. Dispersion distributions are constructed from the data using a kernel density estimator, with optimum bandwidth chosen using Silverman's method. 
In the galaxy as a whole (top panel), average dispersion increases with average age. The B9 region has significantly hotter kinematics than the rest of the galaxy.  \label{fig_sighist}}
\end{figure}

To create 2D dispersion maps, displayed in Figure~\ref{fig_sigmavmap} and Figure~\ref{fig_herschel}, the smoothing radius is fixed to $200''$ for the MS and RGB bins and $275''$ for the less densely populated AGB bins. Stars with fewer than 15 neighbors are dropped from the sample, as their dispersions and associated uncertainties are unreliable. These stars are still available to serve as ``neighbors'' for nearby targets, but the dispersion measurements centered on them are not used. Using smaller smoothing lengths increases spatial resolution in densely packed regions of the survey area, but results in many points being cut because they do not have enough neighbors to yield reliable dispersions and uncertainties. Using larger smoothing lengths further reduces spatial resolution in the maps, reducing the contrast between adjacent high-and low-dispersion patches.

To create the 1D dispersion distributions in Figure~\ref{fig_sighist}, the smoothing radius is chosen independently for each target such that the uncertainty on the velocity dispersion is constant within an age bin. A constant uncertainty makes it easier to understand the spread in the dispersion distribution due to measurement uncertainty. We arbitrarily choose uncertainties of $5~\rm km~s^{-1}$ for the MS+ bin, $7~\rm km~s^{-1}$ for the RGB stars, and $10~\rm km~s^{-1}$ for the AGB stars.  Requiring a constant uncertainty means that stars in regions of lower target density or with larger individual measurement uncertainties have larger smoothing circles.  Again, dispersions measured using fewer than 15 neighbors are dropped. These uncertainty choices result in an average smoothing circle size of about $200''$ for all age bins. If we allow the dispersion uncertainty to be larger, the dispersion distributions are smeared out too much to see individual features. If we cap the uncertainty at a small value, too many points are removed due to small numbers of neighbors.

 It is clear from both the 2D dispersion map in Figure~\ref{fig_sigmavmap} and the 1D dispersion distributions in Figure~\ref{fig_sighist} that the typical
dispersion increases with average age -- that is, older populations
are dynamically hotter than younger populations. In addition, the 2D maps appear to
be patchy, but we will show in \S\,\ref{ssec_structure} that much of
the small-scale spatial variation in dispersion is an effect of finite sampling and is not physical.

%============================================
%============================================

%============================================
%============================================

\subsection{RGB velocity dispersion maps as a function of metallicity}

%============================================
%============================================

We also examine how the RGB dispersion map varies with metallicity.

\begin{figure*}
\scalebox{0.85}{\includegraphics[trim=10 30 15 380, clip =
  true]{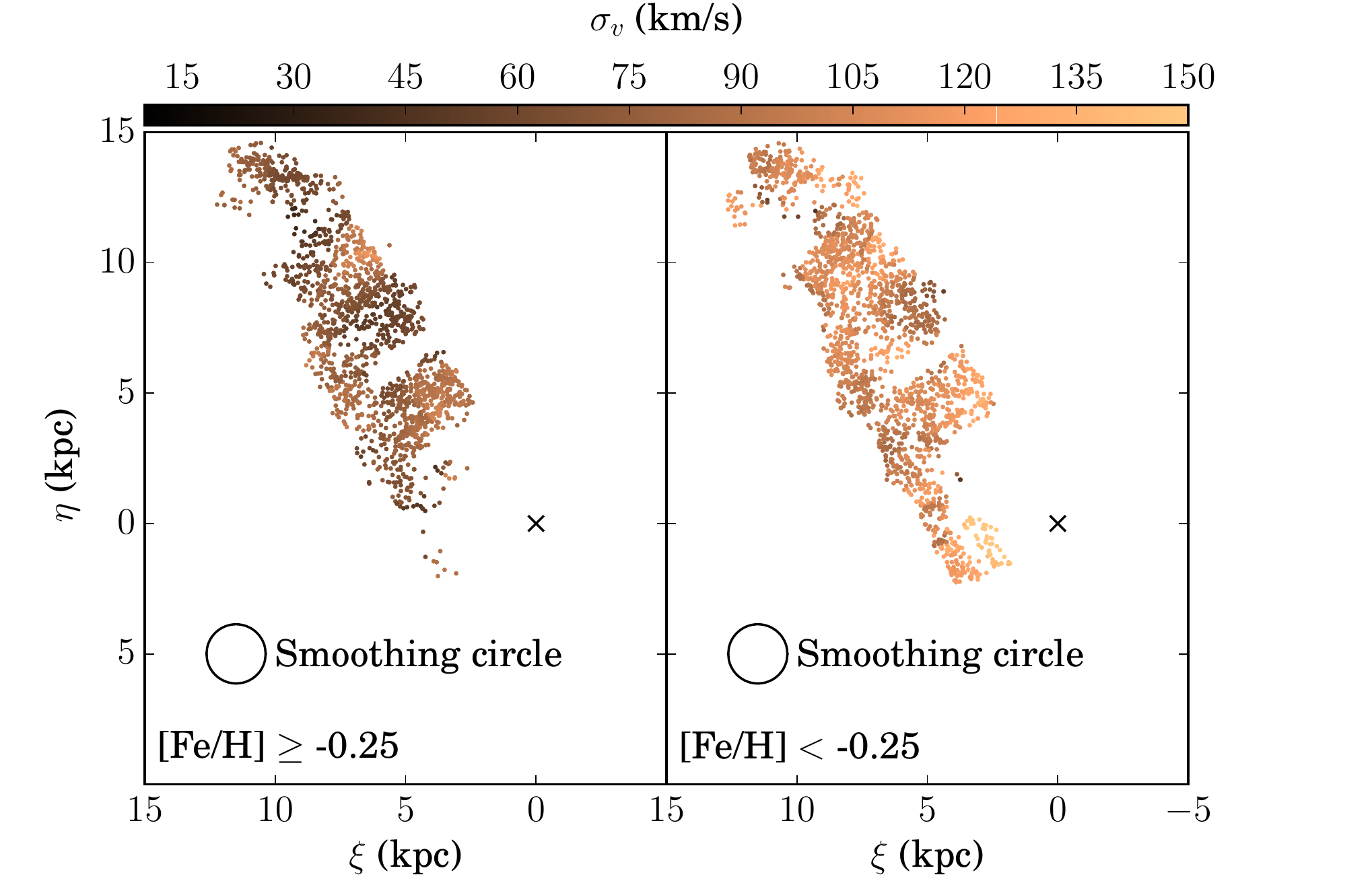}}
\centering
\caption{Smoothed velocity dispersions of RGB stars in high- ({\em
    Left}) and lower-metallicity ({\em Right}) portions of the RGB
  sample. Metallicities are measured using 10 Gyr old \citet{bre12}
  isochrones. On average, the metal-rich population has a lower velocity
  dispersion. Some of the hot patches in the high-metallicity bin may be due to contamination from reddened low-metallicity stars. \label{fig_FeHmap}}
\end{figure*}

\begin{figure}
\scalebox{0.85}{\includegraphics[trim=10 40 200 70, clip =
  true]{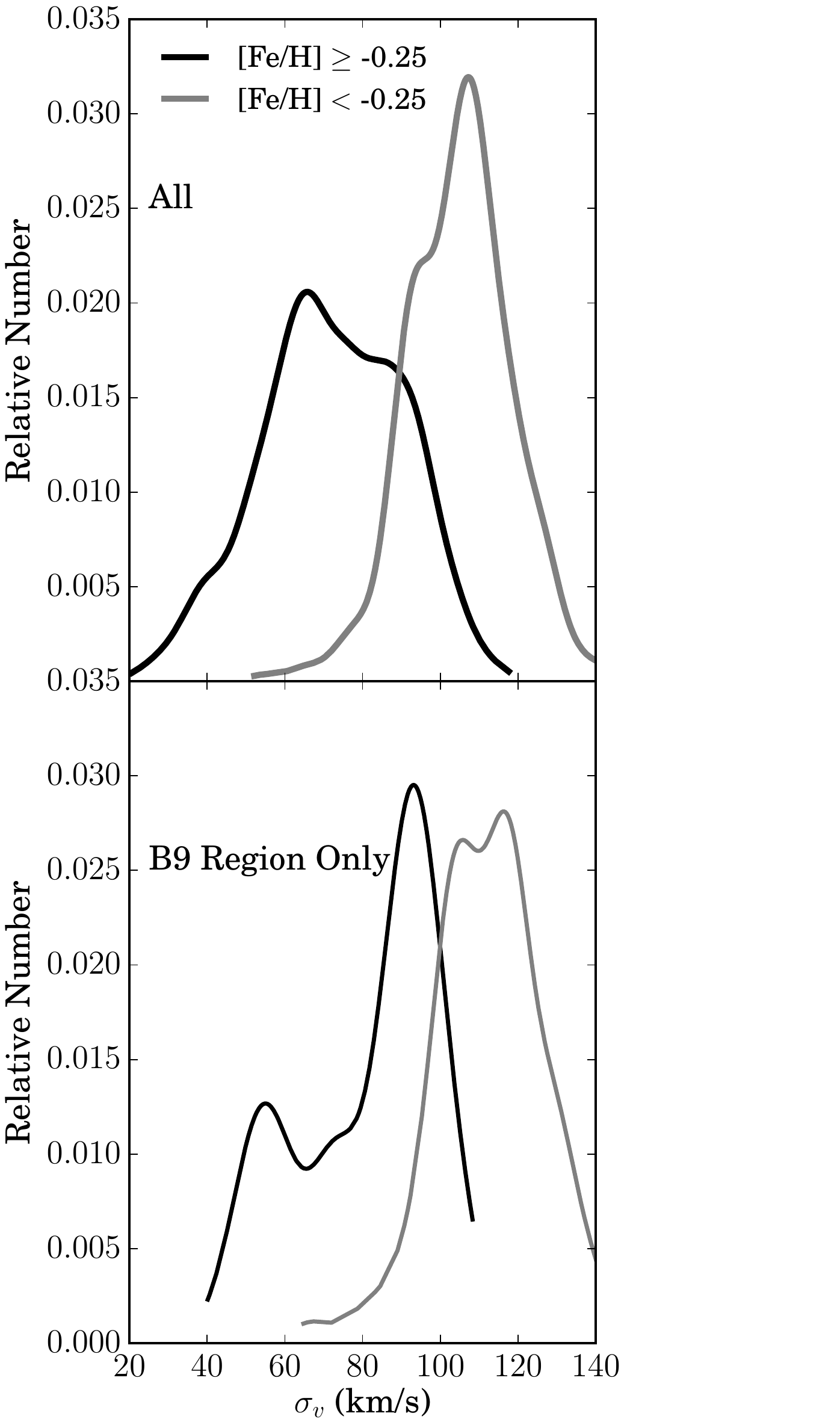}}
\centering
\caption{Same as Figure~\ref{fig_sighist}, for the RGB stars in the two metallicity bins from
  Figure~\ref{fig_FeHmap}. The top panel includes all stars, whereas the bottom panel only includes stars in the Brick 9 region. The metal-rich population is
  dynamically colder on average, and its dispersion is higher on the B9 region than off.  \label{fig_sighistFeH}}
\end{figure}

We estimate the metallicities of the RGB stars by interpolating on the
grid of 10 Gyr-old \citet{bre12} isochrones described in Section 3.1. The
resulting metallicity distribution ranges from $-2.3 < [M/H] < 0.3$
and is strongly skewed towards high metallicities, with a peak at
$[M/H]=-0.2$. We then
split the RGB
sample approximately in half, into a high-metallicity $([M/H] > -0.25$) and
a lower-metallicity bin.  The
stars in our RGB bin are not all 10 Gyr old; the age-metallicity
degeneracy on the RGB means that the metal-poor bin has a slightly higher average age than the metal-rich bin, although there is significant overlap in the age distributions. When we divide the RGB stars in the simulated CMD into
the same two bins by position in the CMD, the low-metallicity bin has an
median age of 2.7 Gyr and a mean age of 4.9 Gyr, with a broad, relatively flat age distribution.  Meanwhile, the high-metallicity bin is slightly younger with a median age of $2.4$ Gyr and a mean age of $3.5$ Gyr. For each bin, we construct a smoothed velocity dispersion map and a
dispersion histogram as described in \S\,\ref{ssec_sv}. The maps are
displayed in Figure~\ref{fig_FeHmap} and the 1D distributions in
Figure~\ref{fig_sighistFeH}. As before, for ease of interpretation, the smoothing circle radius is held constant 
at $200''$ for the maps, and is allowed to vary in order to reach a target dispersion uncertainty of $7~\rm km~s^{-1}$ for the 1D distributions. 

%============================================

%============================================
%============================================

\section{Results}\label{sec_results} 
\subsection{Age-dispersion relation}

\begin{figure}
\scalebox{0.6}{\includegraphics[trim=15 15 15 370, clip =
  true]{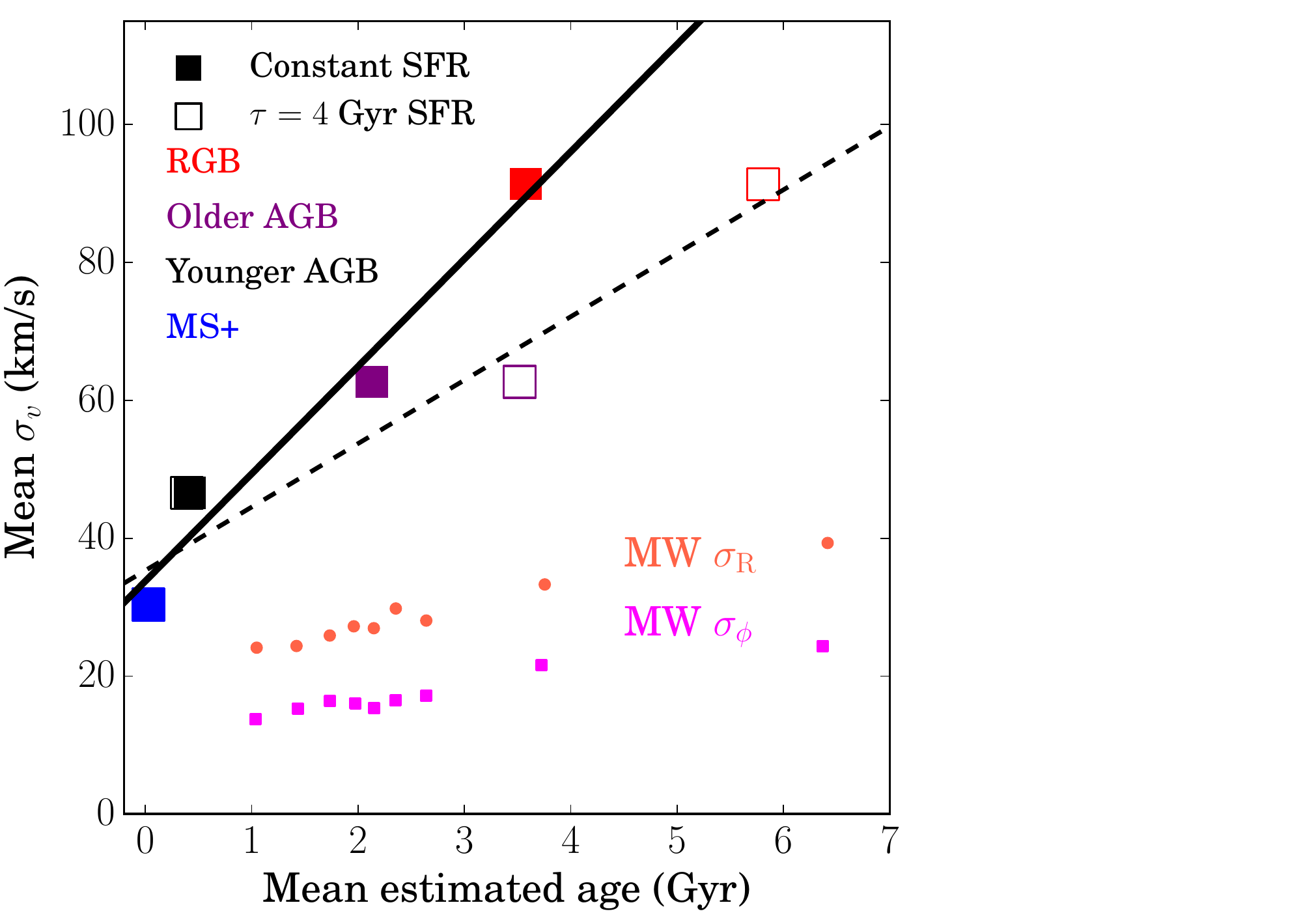}}
\centering
\caption{Comparison of dispersion versus age relations for M31 and the MW solar neighborhood. Large squares: age-dispersion relation for four M31 age bins. Ages are the
  mean ages from the simulated CMD; dispersions are the mean values
  from Figure~\ref{fig_sighist}. Large solid squares and large open squares correspond to ages measured from the constant SFR and decreasing SFR simulated CMDs, respectively. The solid and dashed lines are the best fit
  to the constant SFR and decreasing SFR data points, respectively. Though this is a very rough comparison, the
dispersion of a population appears to be a monotonic function of its
average age. The orange circles and pink squares trace, respectively, the radial and azimuthal dispersion-age profiles for F and G dwarfs from the Geneva-Copenhagen survey \citep{nor04}. Both the slope and normalization of the dispersion versus age relation are much higher in M31 than in the MW,  regardless of the assumed SFR. }\label{fig_heatingrate}
\end{figure}

We can now measure, for the first time, the age-velocity dispersion relation in an external galaxy. Figure~\ref{fig_heatingrate} shows the average velocity dispersion of stars in our four age bins versus their average ages, as estimated from the simulated CMDs. The solid black line shows the best fit to the points assuming a constant SFR, and the dashed black line assuming a decreasing SFR. For comparison, we also show the age-dispersion relation for F and G dwarfs in the solar neighborhood of the Milky Way from \citet{nor04}. We show both the $\sigma_R$ and $\sigma_{\phi}$ components of the Milky Way's velocity ellipsoid. $\sigma_{\phi}$ is the best comparison to our major-axis-dominated M31 data set, but in general the LOS dispersion at any location is some combination of the two components (with a negligible contribution from the vertical component $\sigma_z$). The difference in both slope and normalization is striking. The dispersion of M31 stars increases with age more than 3 times faster than the dispersion of MW stars in the case of the exponentially decreasing SFR, and five times faster in the case of the constant SFR. Additionally, the average dispersion of the RGB bin $(90~\rm km~s^{-1})$ is nearly three times as high as the oldest, hottest population probed in the MW thin disk by \citet{nor04}: $10~\rm Gyr$ stars with $\sigma_{\phi}\sim 32\rm~km~s^{-1}$. The MW's metal-poor thick disk is slightly more disturbed, but still $50\%$ cooler than M31's disk, at $\sigma_R=60~\rm km~s^{-1}$ \citep{bud14}. 

This age-dispersion correlation is robust against contamination of
our age bins. Two bins --- the RGB and older AGB --- are probably
contaminated by a few younger stars. The RGB age bin consists primarily of old
stars, but also includes some higher-mass, intermediate-age stars.
This contamination biases the RGB
dispersion distribution towards that of the faint AGB; in other words,
an exclusively old population would be at least as dynamically hot as
our RGB bin. 
The other affected bin is the older (fainter) AGB group, contaminated by younger AGB stars that have been reddened by dust in the disk of M31.
Again, this contamination biases the faint AGB dispersion
distribution towards smaller values, so that the difference in typical
dispersions between uncontaminated young-intermediate and
older-intermediate age populations is at least as big as the one we
report in Figure~\ref{fig_sighist}.

The high dispersion of M31's stellar disk is also robust against
  ``smearing'' of the LOS component of the rotation velocity $v_{\rm
    Disk LOS}$  within
  the finite-sized smoothing circles. To confirm this, we conduct a test in which we map the 
  dispersion in $(v - v_{\rm Disk LOS})$ instead of just $v$. Here,
  $v_{\rm Disk LOS}$ is computed assuming that the inclination
  $i=77^{\circ}$ and major axis P.A$=38^{\circ}$ over the entire disk,
  and that the rotation velocity of the disk $v_{\rm Disk Rot}$  is a constant within each
  age bin.  $v_{\rm Rot}$ is calculated by fitting to the deprojected
  velocity field of stars in each age bin, and comes out to $(260,
  250, 220)~\rm km~s^{-1}$ for the (MS+, AGB, RGB) stars,
  respectively. The resulting dispersion in $(v - v_{\rm Disk LOS})$
  corresponds to a median correction of just $0.5\%$ for the RGB
  dispersion, $1\%$ for the AGB dispersion, and $6\%$ for the MS+
  dispersion.

We will discuss the evolutionary
implications of the age-dispersion relation in more detail in
\S\,\ref{sec_discussion}. 

%============================================
%============================================

\subsection{Age-metallicity relation among RGBs}

Figures~\ref{fig_FeHmap} and \ref{fig_sighistFeH} show that the metal-poor RGB population
is dynamically hotter than the metal-rich population, by a factor of
almost $50\%$. The metal-rich component has kinematics only about $15\%$ hotter
than the older AGB
population. It is dynamically hot
($\sigma_v \sim 90~\rm km~s^{-1}$) in two spots on the disk.  It should be noted that the metal-rich bin likely contains some reddened metal-poor stars; this may explain the broad dispersion distribution of the metal-rich bin. 

\subsection{Structure in dispersion map}\label{ssec_structure}
The spatial dispersion maps in
Figure~\ref{fig_sigmavmap} show that the the dispersion of all four age bins varies across the face of the disk. 

One of the most obvious features in all four age bins is a
kinematically hot patch about $6~$kpc from the center of the galaxy on
the major axis. This patch is marked by a teal line and outlined in both panels of
Figure~\ref{fig_herschel}. We dub the region enclosed by the black
box the ``Brick 9 region,'' since it is centered on the same area as the set of HST pointings known as ``Brick 9'' in the PHAT survey tiling pattern. The two panels in Figure~\ref{fig_sighist} contrast the dispersion distributions for the full survey and for the brick 9 region. For every age group, the dispersion of stars in the Brick 9 region alone is much higher than for the overall survey. 

 In \S\,\ref{sec_discussion}, we will explore the possible
relationship between this hot patch and M31's bar.

Outside of the Brick 9 region, the dispersion varies with
position (is ``kinematically clumpy''), although the dispersions of
different populations do not necessarily follow the same spatial
pattern. Much of this clumpiness is simply a result of low spectroscopic target density, as we now show using a modified version of the disk toy model described earlier. We construct and analyze the toy model for the RGB disk as an example; similar experiments could be run for the younger stellar populations as well. 

We construct a toy model of a disk by scattering $N$ stars uniformly in a volume $0.8$~kpc thick whose base is a circle of radius $20$~kpc. The velocity vectors of the stars are drawn from a 3-dimensional random normal distribution corresponding to a rotation velocity of $220~\rm km~s^{-1}$ and a velocity dispersion of $90~\rm km~s^{-1}$ (the average observed RGB dispersion) in each direction. (While it is unlikely that the velocity ellipsoid of M31's disk is actually isotropic, this choice does not affect the qualitative results here.) We choose $N$ such that it most accurately reproduces the sampling density of the SPLASH survey.

\begin{figure}
\scalebox{0.8}{\includegraphics[trim= 135 0 0 390, clip =
  true]{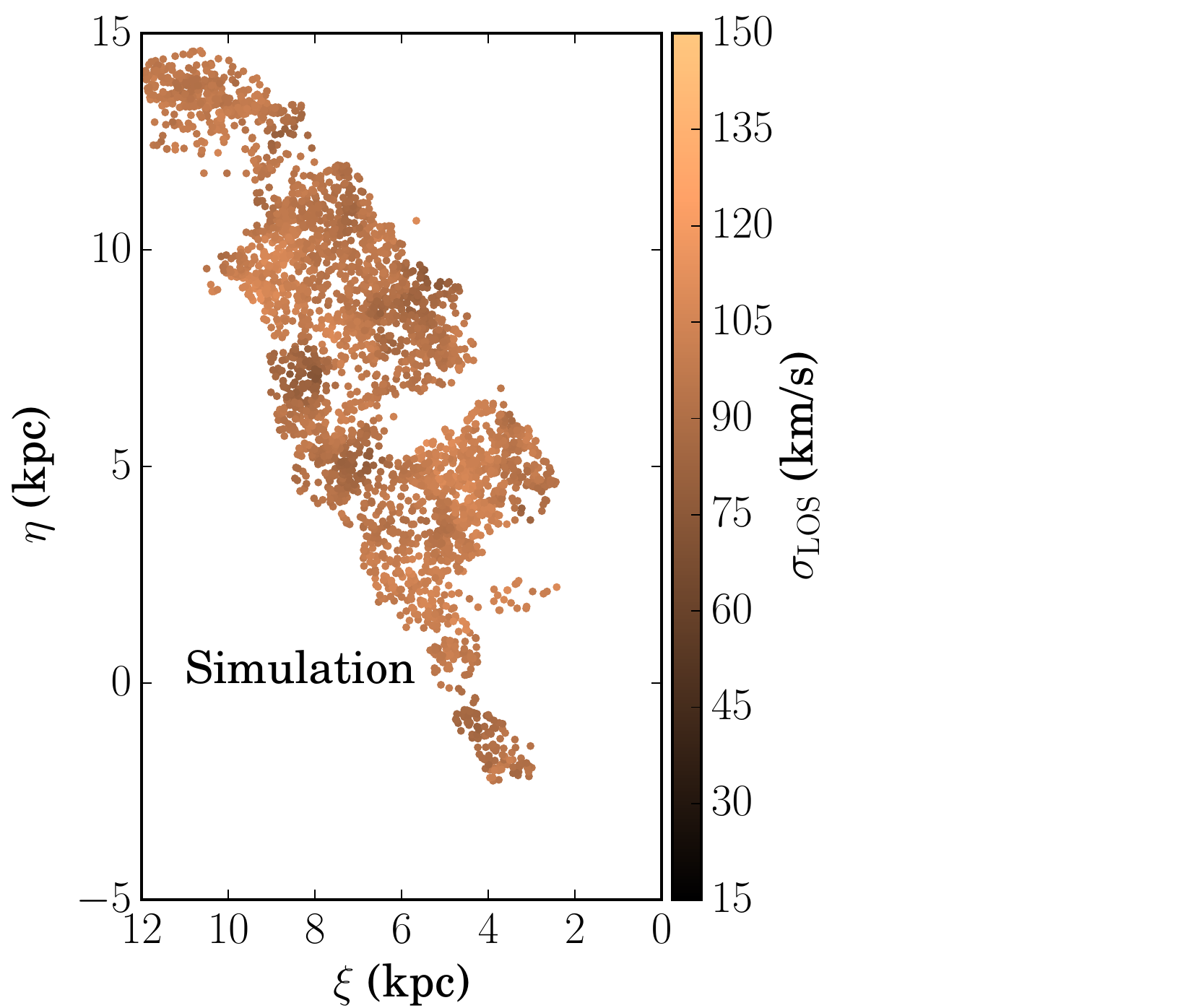}}
\centering
\caption{Velocity dispersion map for a toy model of a disk with a Gaussian velocity distribution with $v_{\phi}=220~\rm km~s^{-1}$ and $\sigma_R = \sigma_{\phi}=90~\rm km~s^{-1}$. \label{fig_simSigmaMap}}
\end{figure}

\begin{figure}
\scalebox{0.8}{\includegraphics[trim=40 0 200 390, clip =
  true]{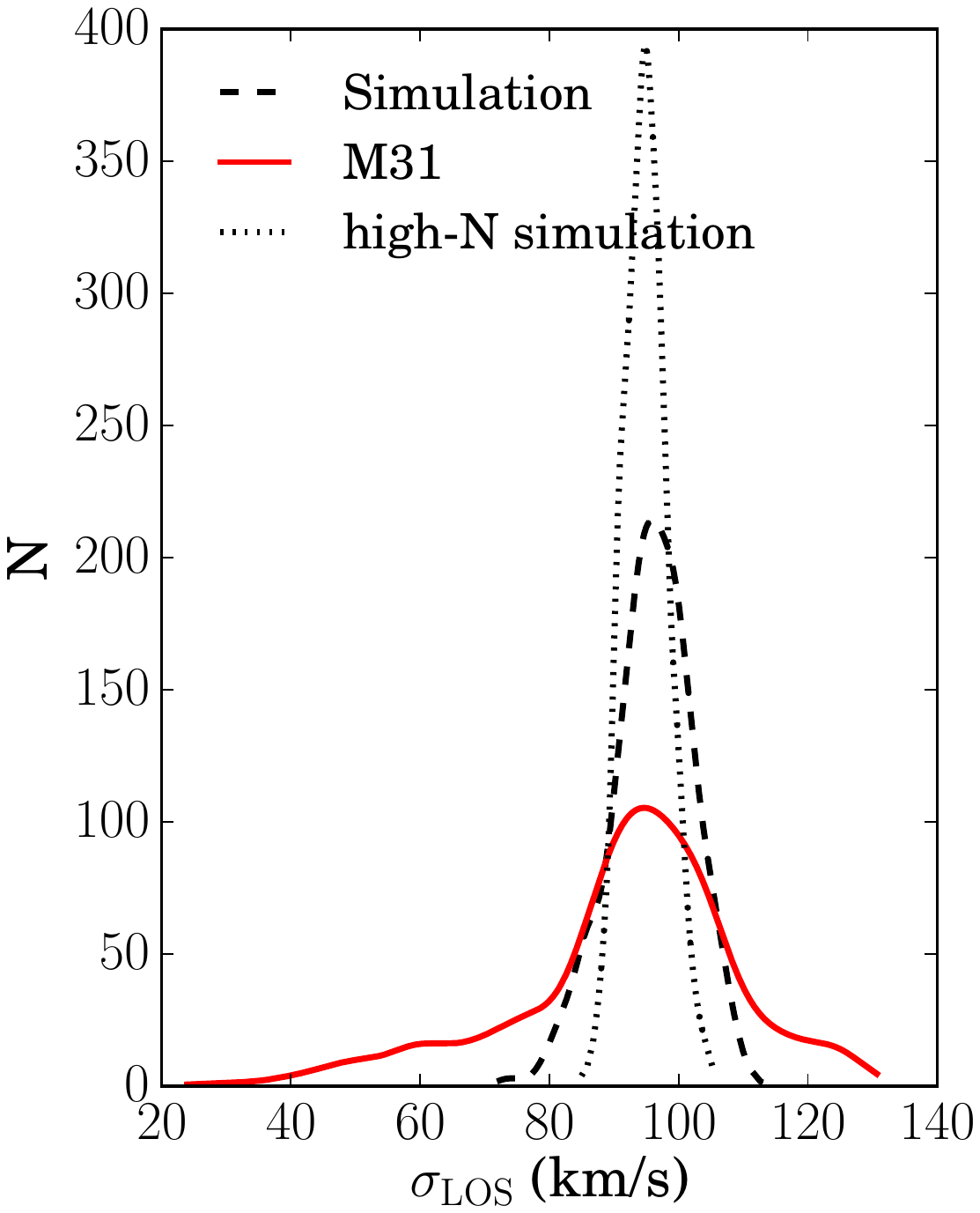}}
\centering
\caption{Velocity dispersion distribution for RGB stars from the toy model of a disk with target density comparable to our spectroscopic survey (black dashed line), from a toy model with target density 5 times higher (blacked dotted line), and from the M31 data (red line). Finite sample density can account for most of the width of the observed dispersion distribution. \label{fig_simSigma1D}}
\end{figure}

We then sample from our toy disk exactly as we do in our spectroscopic sample: at the location of each SPLASH target, we compute the LOS velocity dispersion of all neighbors within a $200''$ radius. The resulting dispersion map, shown in Figure~\ref{fig_simSigmaMap}, is kinematically clumpy. In Figure~\ref{fig_simSigma1D}, we quantify the clumpiness and compare the simulation's 1D dispersion distribution with M31's RGB dispersion distribution. The two distributions have similar full widths at half maximum. To test the effect of sampling density, we also run a simulation with a number density five times higher. The resulting dispersion distribution is only half as wide as the original simulation, corresponding to a less patchy dispersion map.  

Why does finite sampling density cause patchiness in the dispersion map? With few samples, velocity outliers (those on the tails of the velocity distribution) will be scarce and will not fall into every smoothing circle. Velocity dispersion estimates from different smoothing circles may be inflated or deflated based on how many velocity outliers they happen to include. The average dispersion measured accurately reflects the input (intrinsic) velocity distribution ($90~\rm km~s^{-1}$ in this case), but the spread in dispersions from star to star is an artifact of low sampling density.

Figure~\ref{fig_simSigma1D} shows that M31's dispersion distribution has large high- and low-dispersion tails in excess of the simulation --- that is, there are dynamically hot and cold patches that cannot be explained by finite sampling density. This excess clumpiness is not a result of patchy dust: the spatial distributions of dust \citep{dal14} and velocity dispersion are uncorrelated. In \S\,\ref{sec_discussion} we will argue that the dynamically hot patches within each age bin are regions where a few stars from a second kinematical component (for example, a bar, halo, or tidal debris) are overlaid on the uniform underlying disk. 

%============================================
%============================================

%============================================
%============================================
\section{Discussion}\label{sec_discussion}

\subsection{High-dispersion population: superposition of
  two kinematical components}\label{ssec_asym}
  
Thus far, we have shown that the second moment of the local velocity
distribution (LOSVD)  --- the local line-of-sight velocity dispersion ---
varies substantially between age bins and also across the disk of the
galaxy, and that much of the variation across the disk can be explained by finite target density in the spectroscopic survey. Here, we show that the {\em shape} of the LOSVD within each age bin changes with dispersion in a way that cannot be explained by sampling density alone, but can be explained by the presence of a second kinematical component superimposed on a uniform disk. 

If some regions do have intrinsically higher velocity dispersions (not caused by sampling effects), it is not immediately clear whether the
high-dispersion patches contain a single dynamically hotter disk component, or whether
they contain a higher fraction of stars  from a second kinematical
component (such as a spheroid, bar, or tidal stream debris). The two scenarios can be distinguished by the shape of the velocity distribution: If we assume that the
local LOSVD  of the disk is Gaussian (which is a reasonable
approximation in a galaxy with the rotation speed of M31), the former scenario
would produce a wide symmetric Gaussian distribution, whereas the latter would
produce a superposition of two distributions: a narrow central distribution with a wide tail to one side.

Here we test the symmetry of the velocity distribution of low-and high-dispersion patches in the RGB age bin. We repeat the same process for the other three bins.

We first split the
RGB velocity field into nonoverlapping square pixels, $1~\rm kpc$ on
a side. Then, for each pixel, we zero and normalize the velocity
distribution by subtracting the mean and scaling by the standard deviation of velocities within that pixel: $x=(v - \overline{v})/{\rm
  std}(v)$. We now stack all of the scaled velocity distributions $x$. If
the local LOSVD were always Gaussian, the stacked distribution would be a
unit normal symmetric about $x=0$. 

\begin{figure}
\scalebox{0.9}{\includegraphics[trim=25 0 150 390, clip =
  true]{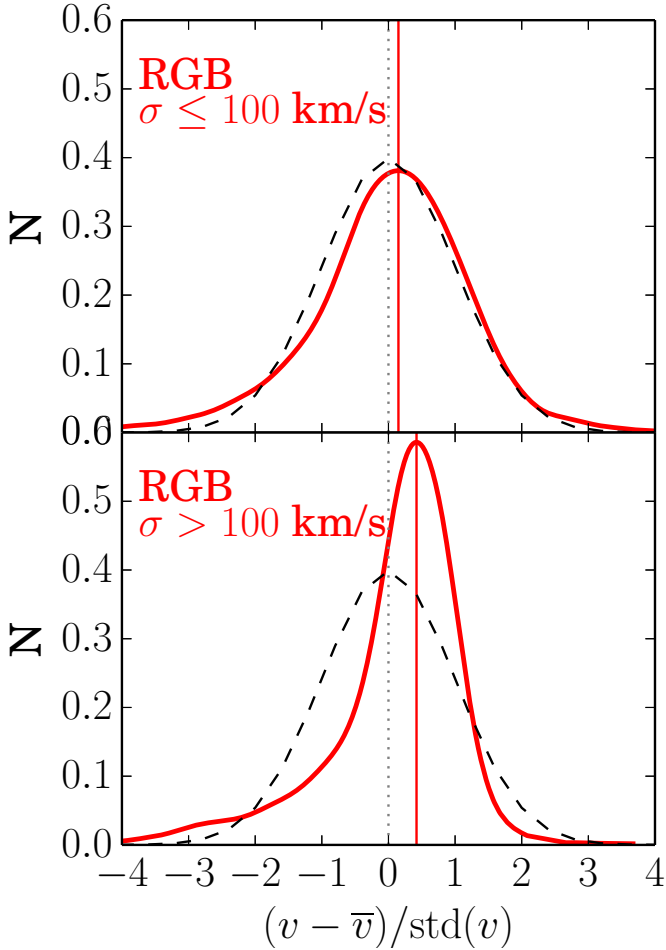}}
\centering
\caption{Comparison of composite velocity distribution shapes for RGB stars
  (solid red curves) to a normal distribution (dashed black
  curves).  The top panel includes pixels with dispersion less than
  $100~\rm km~s^{-1}$, and the bottom panel includes the rest of the
  pixels. By
  construction, all
  distributions have the same mean of 0 and variance of 1. In each panel, the modes of the LOSVD and normal distributions are marked
  by the solid and dotted vertical lines, respectively. The velocity distribution of
  the high dispersion pixels
  (bottom panel)  has a significant low-velocity tail that is barely
  present in the low-dispersion distribution. The inflated dispersion
  of the dynamically hot patches in the RGB dispersion map, then, is due to a
  contribution from a few low-velocity stars in a second kinematical
  component (spheroid or tidal debris), rather than
  to an intrinsically hotter underlying disk.  Though not shown, the same trend is seen
  in the AGB bins, but not the MS bin or in the simulated disk. \label{fig_testgauss}}
\end{figure}

Figure~\ref{fig_testgauss} compares the scaled,
stacked velocity distributions and a unit normal for RGB stars in
pixels with lower dispersions (less than $100~\rm km~s^{-1}$; top panel) and
higher dispersions (bottom panel). {The velocity distributions are skewed to negative velocities, which is towards M31's systemic velocity on this redshifted side of M31's disk.} The stacked distribution composed of the
low-dispersion pixels is nearly symmetric about zero. However, the
high-dispersion velocity distribution is skewed: while its mean is at
$x=0$ by construction, its mode is at a positive velocity
and it has a wide tail towards negative velocities. The same trend is
seen, to a smaller degree, in both the AGB bins, but not in the MS+ bin. 

The skew in the high-dispersion pixels is notably absent when the same analysis is run on the simulated toy disk from \S\,4.1. In that case, both the high- and low-dispersion pixels have symmetric (that is, one-component) LOSVDs. 

In the data, the lower-dispersion pixels always have approximately symmetric velocity distributions, suggestion that they are also dominated by stars from a single kinematical component and that the median dispersions are reliable indicators of the width of the LOSVD of that component. For these lower-dispersion pixels, the correlation between average dispersion and age does indeed reflect a real trend in the degree of heating of the dominant disk. However, within an age bin, the dispersion in dynamically hotter patches are inflated due to a contribution from a few low-velocity stars, rather than to a single, dynamically hot disk component. The spatial variation in dispersion simply reflects inhomogeneity in the number density of this second population. 

The spatially inhomogeneous dynamically hot component has been identified before: first, we identified and mapped the population of stars whose high velocities relative to the local disk LOS velocity rendered them likely "inner spheroid" members \citep{dor12}. Then, we found that there are more kinematically-identified inner spheroid members than can be accounted for by the disk-dominated surface brightness profile, and that the spheroid members have a luminosity function nearly identical to that of the disk \citep{dor13}. The high-dispersion regions identified in this paper are simply regions that include some of these "inner spheroid" stars. The nonuniform spatial distribution of this population could easily be produced by localized effects such as satellite accretion or disk heating via interactions with the bar or with satellites (as proposed in \citet{dor13}) among other things.  As in \citet{dor13}, the hot component is only distinguishable from the rest of the disk via kinematics, not photometry: RGB stars that form the low-velocity tail have the same distribution in color-magnitude space as the rest of the RGB population.

The hot kinematical component inflates the smoothed velocity dispersion in certain pixels above that of the underlying disk, and therefore complicates the comparison to the Milky Way data in Figure~\ref{fig_heatingrate}. The Milky Way data from the Geneva-Copenhagen survey sample only stars within $\sim 40~\rm pc$ of the Sun \citep{nor04}; since the scale height of old stars in the Galaxy's thin disk is several hundred pc \citep[e.g.,][]{bov12}, the GCS data set is strongly dominated by disk stars and includes only minimal contribution from a spatially extended, dynamically hot halo. If the hot population in M31 is much more vertically extended than the underlying colder component, then part of the steep slope of the M31 age-dispersion relation may be due to  "contamination" from a halo-like component that wasn't even probed in the MW survey. However, even in this case, the qualitative point that M31's disk is hotter than the MW's still stands: if we eliminate the likely "hot component" stars (those on the low-velocity tail of the high-dispersion patches in Figure~\ref{fig_testgauss}), the remaining RGB population still has an average dispersion of $80~\rm km~s^{-1}$, twice as high as in the oldest MW stars in the \citet{nor04} sample.
 
%============================================

%============================================

\subsection{Disk evolution scenarios}

In this section, we discuss possible disk formation and evolution scenarios' ability to reproduce the observed features of M31's stellar disk:  a positive correlation between age and velocity
dispersion, a negative correlation between RGB metallicity and
velocity dispersion, a high overall dispersion, and the skewed LOSVD in the high-dispersion patches. 
Dynamically hot populations spatially coincident with a galaxy disk
can arise from three broad categories of processes: dynamical heating
of an initially thin disk; {\em in situ} formation of a thick disk at
high redshift; or accretion of satellite debris onto the disk. In this
section we walk through the three scenarios. We do not attempt to
choose the ``correct'' scenario; that will require more detailed
modeling. However, we show that we can explain all of the observed
features with a combination of some mechanism to create an age-dispersion
relation (either thin disk heating or thick disk collapse) plus some
mechanism to heat the disk nonuniformly (such as accretion of satellites or kicking up of existing
disk stars via satellite impacts).

\subsubsection{Heated thin disk}\label{ssec_heating}

The canonical explanation of thick disk formation involves heating
an initially thin disk via perterbations
from internal structures such as a bar, spiral arms, or giant
molecular clouds (GMCs)
\citep[e.g.,][and references therein]{sel14}; N-body simulations
suggest that impacts from satellite galaxies can also significantly
heat an existing disk \citep{pur09, mcc12, tis13}. If these heating processes happen over an extended period of time,
then the heating a stellar population has undergone should correlate
with its age, consistent with the trend we see in Figures~\ref{fig_sigmavmap}, \ref{fig_sighist},
and \ref{fig_heatingrate}. It is unclear whether internal heating from
GMC scattering or spiral arm perturbations could inflate dispersions
by $60~\rm km~s^{-1}$ over a period of a several Gyr, but stars
kicked out of the disk as a result of satellite impacts can reach 
dispersions of over $100~\rm km~s^{-1}$  \citep{pur10, mcc12, tis13}. 
  
The young stars in our sample have a velocity dispersion $>50\%$ higher than the young stars in the Milky Way \citep{nor04}. If disk evolution is entirely due to heating of an initially thin disk, then M31's disk must have already been born with a higher dispersion. 

Heating from several discrete satellite impacts can explain the skewed LOSVD in patches in the dispersion map.

\subsubsection{Progressive collapse of a thick gas disk}

An alternative to the heated-thin disk scenario described above involves
the collapse of an initially thick gas disk. At early times, a thick, clumpy
disk of gas can form a thick disk
of stars. The remaining gas later collapses into progressively thinner, dynamically colder layers, forming
stars along the way \citep{bou09, for12}. In this way,
younger (more recently formed) stars lie in a colder, more
metal-rich disk than older stars. Such a scenario produces a continuous age-dispersion correlation. Combined with a small amount of heating from satellite encounters and radial migration, it can approximately reproduce both the magnitude and slope of the age-dispersion relationship in the disk of the Milky Way \citep{bir13, bir15}. 

While this disk settling formation mechanism can explain the continuous nature of the age-dispersion correlation in M31 as seen in Figure~\ref{fig_heatingrate}, it cannot reproduce the steepness of the age-dispersion trend $(16~\rm km~s^{-1} Gyr^{-1}$ for the constant SFR or $9$ for the declining $\tau=4~\rm Gyr$ SFR) and the very high velocity dispersion. The
fiducial model in \citet{for12} shows that, at $3$ radial scale
lengths $(R\sim15$ kpc in M31), the dispersion in a MW-mass galaxy 
with a decreasing gas accretion rate and a standard radial migration prescription
increases only $7~\rm
km~s^{-1} ~Gyr^{-1}$ from $0$ to $2$ Gyr --- half that of M31 assuming a constant SFR, and still $20\%$ less than in M31 in the case of the heavily old star-weighted declining SFR. 

Similarly, additional heating is need to explain the extremely high velocity dispersion, which, for the RGB stars,  is $40\%$ of the circular
velocity of $250~\rm km~s^{-1}$ \citep{che09, cor10}. In contrast, the
highest stellar dispersion in models presented in \citet{for12} is
$40~\rm km~s^{-1}$, or $<20\%$ of the circular velocity. 

This scenario also explains the inverse correlation between
metallicity and velocity dispersion for the RGB stars, assuming that
the more metal-rich RGB population is the younger one. 

%This scenario explains the {\em direction} of our
%age-velocity dispersion trend in Figure~\ref{fig_heatingrate}, although the {\em slope} of
%the trend $(16~\rm km~s^{-1} Gyr^{-1}$ for the constant SFR or $9$ for the declining $\tau=4~\rm Gyr$ SFR) would require an extraordinary
%amount of stellar heating via radial migration combined with a very
%small, if any, amount of gas accretion after the initial accretion event. 
\subsubsection{Satellite accretion}
Cosmological simulations predict that stellar halos are built up via
hierarchical merging with smaller galaxies, and the prominent
substructure in the outskirts of the Milky Way \citep[e.g.,][]{mat74,
 yan03, roc03, iba94, maj03, new03} and M31 \citep{iba04, iba07,
 fer02, kal06,gil09, mcc09, tan10, gil12} lend
support to this idea.  Because the total area of the disk is
smaller than that of the extended halo, satellite-disk
interactions should be rarer than satellite-halo interactions, but not nonexistent. In \S\,\ref{ssec_heating}, we discussed how these interactions can heat the existing disk. But stars from the satellite galaxies can also accrete onto the disk, changing the age and metallicity distributions of the disk in addition to dynamically heating it. Debris from at least one
tidal disruption event has been found across the face of M31. The
Giant Southern Stream, Northeast Shelf, and
Western Shelf all show up as kinematically cold substructure
\citep{iba01, kal06, gil09, dor12}. Accretion from satellites can naturally explain the asymmetric velocity distributions in our RGB population. Since
the velocity of the satellite would likely be quite
different than the rotation velocity of the disk, regions with
accreted stars would end up with inflated line-of-sight velocity
dispersions, producing localized structure. However, accretion alone cannot
account for the continuous increase in dispersion with age. It also cannot account for the high dispersion of the young MS+ stars relative to the young stars in the MW unless M31's young stars were born with a higher dispersion.

\subsection{Brick 9 Feature: Associated with the Bar?}

Many of the high-dispersion measurements lie in 
 the ``Brick 9 Region" centered at $(\xi, \eta)=(4.5, 4.5)$ kpc, marked by the ends of the teal line in
 Figure~\ref{fig_herschel}. This high-dispersion region was identified in RGB stars in \citet{dor12} and \citet{dor13} as having the highest fraction of ``inner spheroid" or ``kicked-up disk" stars. Here, we see that it is not limited to the RGB sample. The Brick 9 feature is present in all age bins and in both
RGB metallicity bins, suggesting that it is not an accreted population or formed by an event that ended more than a Gyr ago. It
is located on the major axis of the galaxy, so is not subject to the geometric
dispersion inflation that happens near the minor axis. It does
happen to be located on the inner ring (visible in the Herschel image
in the lower right panel of Figure~\ref{fig_sigmavmap}) and may be
coincident with the end of a long bar. 

\citet{ath06} compared the infrared
isodensity contours and radial luminosity profiles to N-body
simulations of barred galaxies to show that M31 likely has a bar that
may extend out to $1300''~(5~\rm kpc)$ at a position angle of
$45^{\circ}$ --- placing its NE end just interior to the Brick 9
feature. The effect of bars on stellar kinematics is not well
understood, but \citet{ath06} show that the presence of a bar, when
misaligned with the major axis of its host disk, can result in a broad
velocity distribution near the end of the bar. Typically, bar members
have different orbits from disk members, and so smoothing circles containing both
bar and disk members would contain a large spread in line-of-sight
component of stellar velocities. 

\subsection{Comparison to Milky Way} 
The work presented in this paper shows that M31's stellar disk is dynamically
hotter than the MW's: the average dispersion of the RGB bin $(90~\rm km~s^{-1})$ is nearly {\em three times} as high as the oldest, hottest population probed in the MW by \citet{nor04}, and the average dispersion of our youngest age bin is more than $50\%$ higher than that in the Milky Way.  In addition, the slope of the dispersion versus age relation is more than $3$ times higher in M31 than our own galaxy. Both suggest that M31 has had a more violent accretion history than the MW in the recent past.

This result is encouraging for $\Lambda$CDM cosmology, which predicts
that $70\%$ of disks the size of those in the MW and M31 should have
interacted with at least one satellite of mass $\sim  3M_{\rm disk}$
in the last $10~\rm Gyr$  \citep{ste08}. The Milky Way's disk is far too
cold for it to have undergone such an encounter \citep{pur10}, but the
fact that M31's disk is dynamically warmer leaves open the possibility
that $\Lambda$CDM predictions are correct and that the MW is simply an
outlier with an unusually quiescent history. 

 %Both galaxies host a single large tidal stream that dominates the
 %tidal debris at small radii: the Giant Southern Stream (GSS) in M31
 %\citep[e.g.,][]{iba01, guh05, kal06, gil09}, whose progenitor likely
 %had a mass of $\sim10^9 M_{\odot}$, and the Sagittarius stream in
 %the Milky Way \citep[e.g.,][]{iba94, maj03, bel06}, whose progenitor
 %had a similar mass \citep{nie10}. However, 

Our results support a growing body of evidence that M31 has
  experienced a more violent merger history than our own galaxy. M31's
  halo is littered with substructure that can be identified both in photometry \citep{iba14} and kinematics \citep{cha08, gil12}, and fields at similar radii can host clearly distinct stellar populations, indicating that not all stars share a common origin \citep{fer02, ric08}. The number of giant streams discovered in M31 outnumbers of that of the MW \citep{iba07}.  Furthermore, the slope of the surface brightness profile of M31's outer halo is shallow, with projected powerlaw slope $\alpha\sim -2$ \citep{gil12, dor13, iba14}. In comparison, the surface brightness of the MW halo exterior to $50~\rm kpc$ as traced by BHB stars is much steeper, falling as $r^{-6}$, suggesting a more quiescent recent accretion history \citep{dea14}. M31's gas disk is also strongly warped in the outer regions: the major axis position angle of the gas disk changes by $10^{\circ}$, and the  inclination by $7^{\circ}$, between $R=20~{\rm and}~40~\rm kpc$ \citep{che09, cor10}. 

A quantitative comparison between the velocity dispersion of M31's old
stellar disk and cosmological predictions requires the measurement of
the shape of M31's velocity ellipsoid. We save this measurement for a
future paper (Dorman et~al. 2014b, in preparation) because it requires
the choice of a particular disk rotation curve model.

\subsection{Radial trends}

\begin{figure}
\scalebox{0.75}{\includegraphics[trim=15 15 0 00, clip =
  true]{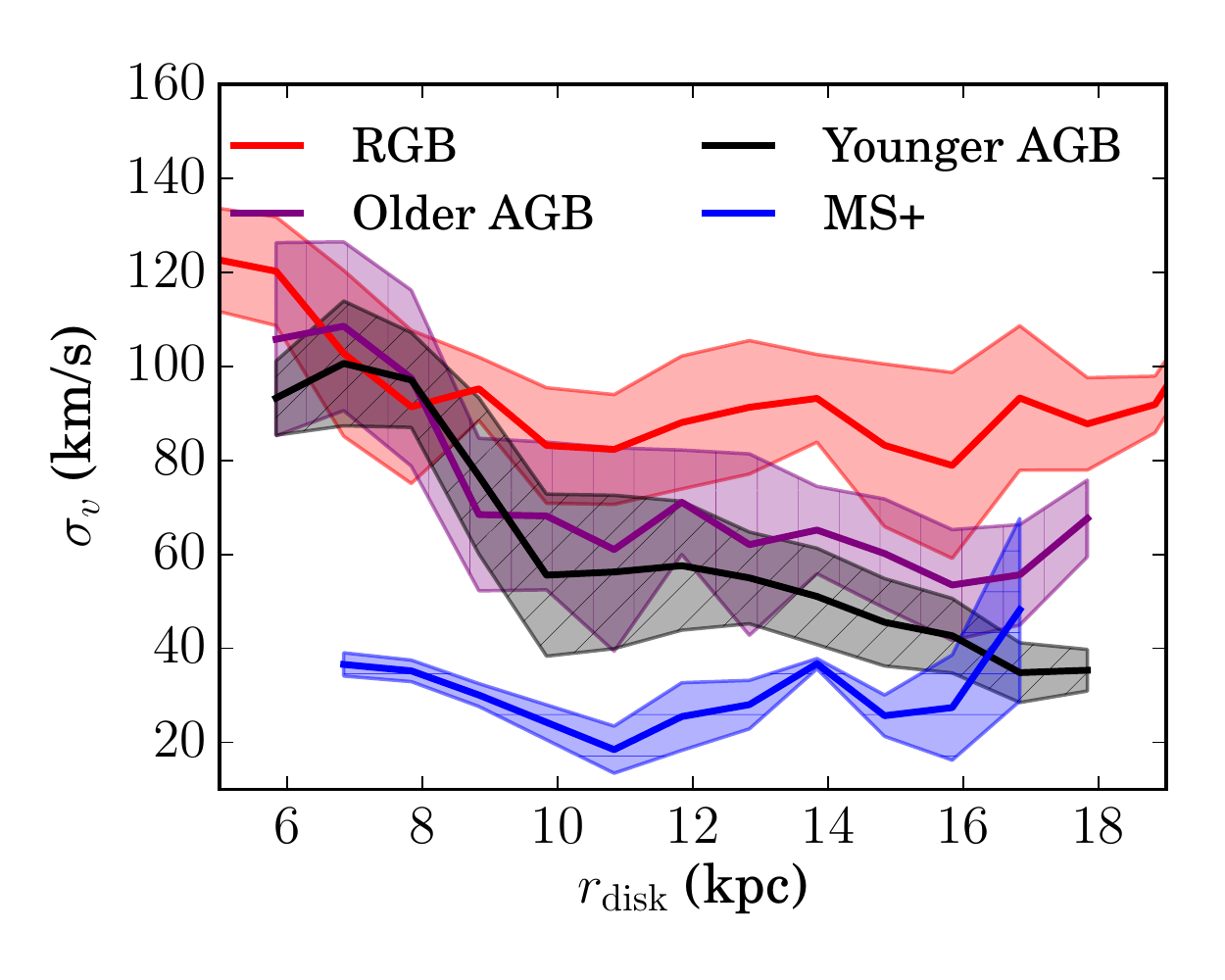}}
\centering
\caption{ Dispersion as a function of radius in the plane of the disk
  $r_{\rm disk}$ for each age bin. Age
  bins are color coded as in previous plots. The solid line and shaded region correspond to the
  median and standard deviation of the velocity dispersion in each 1
  kpc wide radial bin. $r_{\rm disk}$ is
  computed assuming a constant major axis position angle of
  $38^{\circ}$ and disk inclination of $77^{\circ}$. Only stars
  within $\pm 20^{\circ}$ of the major axis are used, since near the
  minor axis distance measurements become more sensitive to
  precision in inclination, and a single smoothing circle may cover a wide
  range of radii. Velocity dispersion decreases with radius in all
  bins out to about $10~\rm kpc$. \label{fig_radial}}
\end{figure}

In Figure~\ref{fig_radial}, we show velocity dispersion as a function
of radius: the median and standard deviation
of the velocity dispersion in each $1~\rm kpc$ wide radial bin in the
plane of the disk for each age bin. We only include stars within $\pm
20^{\circ}$ of the major axis, because near the minor axis a single
smoothing circle over which dispersion is calculated may cover a wide
range of radii. 

 The figure reveals two distinct radial regions in the galaxy.  Inside $10~\rm kpc$, the velocity dispersions of all age groups are inflated by a factor of $1.5-2$ from their values outside, and the dispersions of the three older age groups are similar to each other, especially interior to $8~\rm kpc$. The dispersions of all age groups decline steeply with radius in this inner region. Outside the B9 region, the velocity dispersions of the different age groups separate from one another, revealing the clean age vs. dispersion relationship, and decline less steeply with radius.  The transition radius may correspond to a transition in the dominant heating mechanism: in the outer regions, the clean age-dispersion relation implies a continuous heating process over the past several Gyr. Inside, a single perturbation has modified the orbits of stars older than $100~\rm Myr$. This could be a single event such as a very recent satellite impact, or a large fraction of the older stars in this region could belong to a single, kinematically distinct component such as the bar or a thick disk with short scale length.

The $10~\rm kpc$ break corresponds to a bit less than two radial scale lengths \citep{wor05, sei08, iba05, dor13}. This is consistent with the ``mass follows light" radial trend seen in stellar kinematics in the DiskMass survey of nearly face-on, late-type spiral galaxies \citep{mar13}: the LOS stellar velocity dispersion typically decreases with radius out to two scale lengths, after which the data become noisy enough that the trend becomes consistent with flat. It is worth noting that because the DiskMass survey galaxies are much closer to face-on than Andromeda is $(< 30 ^{\circ} {\rm~vs.~} 77^{\circ})$, the DiskMass LOS dispersion profiles are sensitive to the vertical component of the velocity dispersion $\sigma_Z$ while our LOS velocity dispersion has a negligible contribution from $\sigma_Z$. 

\section{Summary}\label{sec_summary}

We have split a spectroscopic sample of 5800 stars from the disk of M31 into four age groups based on optical HST photometry, and mapped the line-of-sight velocity dispersion of each age group. We have found: 

\begin{itemize}

\item Stellar velocity dispersion and age are directly correlated in the disk of M31. The slope of the best fit line assuming a constant SFR is approximately $16~\rm km~s^{-1}~Gyr^{-1}$, more than $5$ times higher than that of the Milky Way.  

\item Among RGB stars in our sample, the metal-poor half is $50\%$ kinematically hotter than the metal-rich half. 

\item The stellar disk is kinematically clumpy on scales smaller than
  about $200''~(760~\rm pc)$. Most of the structure can be attributed to sampling effects. However, the line of sight velocity distribution
  of the RGB and AGB stars in the highest-dispersion regions is quite
  asymmetric, with a low-velocity tail. This asymmetry indicates the presence of an additional kinematic component that is not smoothly distributed across the galaxy.

\item There is a patch on the major axis about $6$ kpc from the galactic center where the dispersions of all four age groups are inflated by a factor of $1.5-2$. This patch may correspond to the end of the long bar. 

\end{itemize}

 %  ACKNOWLEDGEMENTS

PG and CD acknowledge NSF grants AST-1010039 and AST-1412648 and NASA
grant HST-GO-12055. CD was supported by a NSF Graduate Research
Fellowship and a Eugene Cota-Robles Graduate Fellowship. DRW is supported by NASA through Hubble Fellowship grant HST-HF-51331.01 awarded by the Space Telescope Science Institute. We thank Emily Cunningham, John Forbes, Leo Girardi, Dylan Gregersen, Cliff Johnson, Connie Rockosi, and Graeme Smith for helpful conversations. Finally, we are grateful to the anonymous referee for constructive suggestions that greatly improved the clarity of the paper. 
We acknowledge the very significant
cultural role and reverence that the summit of Mauna Kea has always
had within the indigenous Hawaiian community. We are most fortunate to
have the opportunity to conduct observations from this mountain. 

%\clearpage

\begin{appendix}

This paper presents, for the first time, resolved stellar kinematics
derived from Keck/DEIMOS observations from Fall
2012. Table~\ref{tab_obs} lists the slitmasks observed and the number
of quality radial velocities recovered from each. 

The target selection was done in a non-uniform way in order to
maximize the number of younger (main-sequence, asymptotic giant
branch, blue and red supergiant) stars relative to the dominant red
giant branch population. As such, the spectroscopic selection function varies with
color, magnitude, and position on the sky, though it can be easily
recovered by comparing the distribution of targets to the full PHAT
catalog. 

The raw data were flat-fielded, locally sky subtracted, collapsed into
1-D spectra, and cross-correlated against template spectra to measure
radial velocities as described in \citet{dor12,dor13}. 

%TABLE 1
\begin{deluxetable*}{llllrccccc}[h!]
\tabletypesize{\scriptsize}
\tablecaption{Keck/DEIMOS Multiobject Slitmask Exposures from Fall 2012}
\tablewidth{0 pt}
\tablehead{
\colhead{Mask}&
\colhead{Observation}&
\colhead{$\alpha$ [J2000]}&
\colhead{$\delta$ [J2000]}&
\colhead{P.A.}&
\colhead{$t_{\rm{exp}}$}&
\colhead{Seeing}&
\colhead{No. of}&
\colhead{No. of Usable}&
\colhead{No. of Usable}\\
\colhead{Name}&
\colhead{Date (UT)}&
\colhead{(h m s)}&
\colhead{($^\circ$~$'$~$''$)}&
\colhead{($^\circ$)}&
\colhead{(sec)}&
\colhead{FWHM}&
\colhead{Slits}&
\colhead{Target Velocities}&
\colhead{Velocities of}\\
\colhead{}&
\colhead{}&
\colhead{}&
\colhead{}&
\colhead{}&
\colhead{}&
\colhead{}&
\colhead{}&
\colhead{(Success Rate)}&
\colhead{Serendipitously Detected}\\
\colhead{}&
\colhead{}&
\colhead{}&
\colhead{}&
\colhead{}&
\colhead{}&
\colhead{}&
\colhead{}&
\colhead{}&
\colhead{Stars}
}
%Slit counts: without alignment or guide stars
%Star counts: including zq = 1, 3, 4
\startdata
        mct6C & 2012 Sept 18 & 00 44 47.17 & 41 22 00.0 & -140.0 &
        $3\times1020$ & 0''.6 & 225 & 184 (82\%) & 31\\
        mct6D & 2012 Sept 20 & 00 44 34.60 & 41 29 44.6 & 170.0 &
        $3\times1200$ & 0''.8 & 208 & 167 (80\%) & 38\\
        mct6E & 2012 Sept 18 & 00 44 13.52 & 41 19 05.5 & ~-20.0 &
        $2\times1080$ & 0''.6  & 228 & 107 (47\%) & 16 \\
        mct6F & 2012 Sept 19 & 00 45 54.72 & 41 41 58.6 & -164.0 &
        $3\times1020$ & 0''.6 & 221 & 192 (87\%) & 31 \\
        mct6G & 2012 Sept 20 & 00 45 38.34 & 41 43 37.4 & -155.0 &
        $3\times1200$ & 0''.9 & 231 & 178 (77\%) & 24 \\
        mct6H & 2012 Sept 18 & 00 45 26.92 & 41 44 04.3 & ~+15.0 &
        $3\times1020$ & 0''.6 & 244 & 185 (76\%) & 24 \\
        mct6I & 2012 Sept 19 & 00 44 10.25 & 41 25 16.2 & ~-95.0 &
        $3\times1080$ & 0''.7 & 209  & 138 (66\%) & 52 \\
        mct6K & 2012 Sept 18 & 00 44 38.26 & 41 37 22.5 & ~-95.0 &
        $3\times1020$ & 0''.7 & 207  & 155 (75\%) & 45 \\
        mct6L & 2012 Sept 19 & 00 46 05.81 & 42 02 28.1 & ~-20.0 &
        $3\times1020$ & 0''.8 & 246  & 212 (86\%) & 8 \\
        mct6M & 2012 Sept 20 & 00 44 36.98 & 41 32 39.5 & ~-20.0 &
        $3\times1080$ & 0''.75 &  227  & 153 (67\%) & 40 \\
        mct6O & 2012 Sept 20 & 00 45 08.17 & 41 52 34.0 & ~-80.0 &
        $2\times 1080+1\times 1140$ & 0''.93 & 240  & 169 (70\%) & 24 \\
        mct6P & 2012 Sept 19 & 00 45 30.50 & 41 55 37.6 & ~-80.0 &
        $2\times 1019 + 1\times 855$ & 0''.6 & 229  & 161 (70\%) & 10 \\
        mct6Q & 2012 Sept 18 & 00 45 35.98 & 42 00 17.9 &  ~-80.0 &
        $3\times1080$ & 0''.8 & 225  & 155 (69\%) & 21 \\
        mct6R & 2012 Sept 19 & 00 47 02.28 & 42 12 07.3 &  ~-80.0&
        $3\times1080$ & 0''.8 & 227  & 175 (77\%) & 9 \\
        mct6S & 2012 Sept 20 & 00 47 02.27 & 42 09 25.2 &  ~-80.0 &
        $3\times1200$ & 0''.8 &224  & 163 (73\%) & 18 \\
        mct6T & 2012 Sept 18 & 00 45 57.60 & 42 01 12.0 &  ~-40.0 &
        $3\times1020$ & 0''.6  &245  & 210 (86\%) & 10 \\
        mct6U & 2012 Sept 20 & 00 46 13.24 & 42 14 35.6 & ~+50.0 &
        $3\times1200$ & 0''.8 & 254  & 185 (73\%) & 5 \\
        mct6V & 2012 Sept 18 & 00 46 40.80 & 42 13 48.0 & -130.0 &
        $2\times1080$ + $1\times 1020$ & 0''.6  & 226  & 126 (56\%) & 13 \\
        mct6W & 2012 Sept 19 & 00 46 51.73 &42 12 46.0 & -130.0 &
        $3\times1020$ & 0''.6  & 225 & 175 (78\%) & 8 \\
        mct6X & 2012 Sept 19 & 00 46 22.94 & 42 01 58.0 & 140.0 &
        $3\times1020$ & 0''.7  & 225  & 173 (77\%) & 15 \\

\tableline\tableline \vspace{1 mm}
        \\
	Total: &&&&&&& 4566~ & 3368 (74\%)~ & 442
\enddata

\label{tab_obs}
\end{deluxetable*}

\end{appendix}
\end{document}